\documentclass[manuscript,screen]{acmart}
\usepackage{subfig}
\AtBeginDocument{%
  }

\setcopyright{acmlicensed}
\copyrightyear{2018}
\acmYear{2018}
\acmDOI{XXXXXXX.XXXXXXX}
\acmConference[Conference acronym 'XX]{Make sure to enter the correct
  conference title from your rights confirmation email}{June 03--05,
  2018}{Woodstock, NY}
\acmISBN{978-1-4503-XXXX-X/2018/06}

\begin{document}

\title{Characterizing Memory Misalignment in Human-LLM Interaction From User Perspectives}

\author{Jingruo Chen}
\email{jc3564@cornell.edu}
\affiliation{
    \institution{Information Science, Cornell University}
    \city{Ithaca}
    \state{New York}
    \country{USA}
}

\author{Shuning Zhang}
\email{zsn23@mails.tsinghua.edu.cn}
\affiliation{
    \institution{Tsinghua University}
    \city{Beijing}
    \country{China}
}

\author{Eryue Xu}
\email{eryuexu2@illinois.edu}
\affiliation{
    \institution{School of Information Sciences, University of Illinois Urbana-Champaign}
    \city{Urbana}
    \state{Illinois}
    \country{USA}
}

\author{Jianing Li}
\email{ge85gen@tum.de}
\affiliation{
    \institution{School of Computation, Information and Technology, Technical University of Munich}
    \city{Munich}
    \country{Germany}
}

\author{Xin Yi}
\email{yixin@tsinghua.edu.cn}
\affiliation{
    \institution{Tsinghua University}
    \city{Beijing}
    \country{China}
}

\renewcommand{\shortauthors}{Chen et al.}

\begin{abstract}
   While memory enhances personalization in LLM-based conversational agents, it suffers from memory misalignment, where memories violate user expectations. We present a mixed-methods investigation to characterize and mitigate memory misalignment from user perspectives. First, we collected data from memory usage (N=28, 457 entries) and diary study (N=32, 304 reports), which yielded a taxonomy spanning 14 misalignment types across memory intake, storage and management, retrieval and interpretation stages. Second, four co-design workshops with 12 experienced HCI researchers derived a design space to tackle memory misalignment issues, consisting of 12 candidate interaction strategies structured across interaction form, placement and intrusiveness dimensions. Finally, a speed dating with 121 users reveals preference heterogeneity, where users prioritize proactive controls over cognitively demanding causal graph inspections or passive audit logs. Synthesizing these findings, we highlight the tension between supervisory agency and interaction overhead, and advocate for friction-aware memories that balance user oversight with conversation smoothness.
\end{abstract}





\begin{teaserfigure}
 \includegraphics[width=\textwidth]{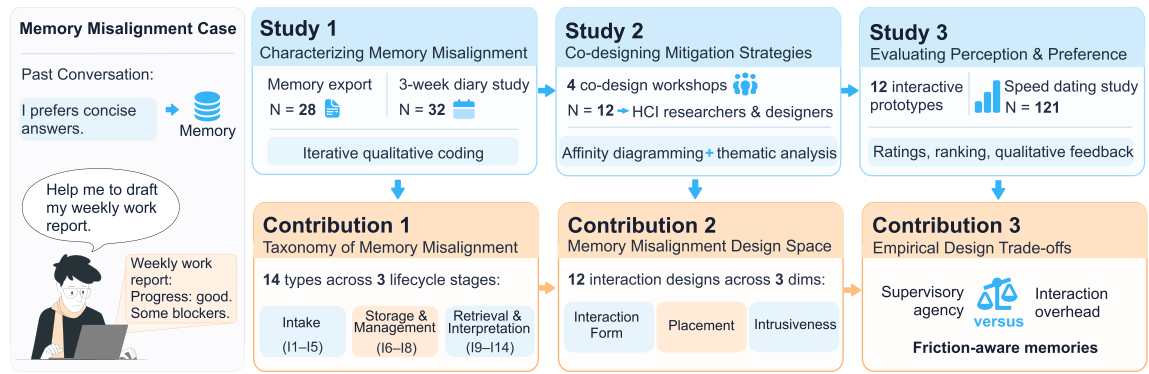}
  \caption{The framework of this paper.}
  \Description{The framework of this paper, which revolves around memory misalignment issues. An example on the left illustrates how information from a past conversation can be stored as memory and affect a later interaction. Three studies then characterize memory misalignment, co-design mitigation strategies, and evaluate the resulting designs. These studies contribute a taxonomy of memory misalignment, a design space for mitigation strategies, and empirical evidence about trade-offs between supervisory agency and interaction overhead.
}
  \label{fig:teaser}
\end{teaserfigure}


\maketitle

\section{Introduction}

Memories are essential for enabling effective long-term interactions in LLM-based conversational agents~\cite{wu2025long}. By retaining contextual cues across sessions, agent memory reduces user effort~\cite{ning2025user} and enables adaptive personalization aligned with user preferences~\cite{ren2019lifelong}. As a result, memory mechanisms have been widely deployed in commercial conversational systems, such as ChatGPT~\cite{chatgpt2026}, Gemini~\cite{gemini2026}, and Doubao~\cite{doubao2026}.

Despite these advancements, deployed memory systems often induce negative user experiences~\cite{pataranutaporn2025slip,xiong2026memory}. Beyond factual errors and hallucinations~\cite{chen2025halumem}, even factually accurate memories could be inappropriately accumulated, retained, or used in situated interactions\footnote{https://www.facebook.com/groups/claudeaicommunity/posts/1356633023170528/}. While recent research has benchmarked factual inaccuracies~\cite{uddin2026recall,xiong2026memory,zhang2024ghost} and general conversational memory retrieval~\cite{bian2026realmem}, little empirical work has investigated \textbf{memory misalignment, the discrepancy between system memory behaviors and user intents or expectations,} from the user's perspective. Understanding these breakdown patterns and designing coping mechanisms is important for developing reliable LLM memory architectures. To address this gap, we propose the following research questions (RQs):

$\bullet$ RQ1: What are the memory misalignment cases in interactions between humans and LLM-based chatbots?

$\bullet$ RQ2: What are interaction designs that can mitigate memory misalignment?

$\bullet$ RQ3: How do users perceive the efficacy of these mitigation strategies?
% effective are these designs from the perspectives of users' perceptions?

For RQ1, we conducted a two-phase data collection study combining a memory database (N=28, 457 memories) and a three-week diary study (N=32, 304 reports). Grounded in cognitive memory error frameworks~\cite{schacter1999seven}, we identified 14 distinct misalignment types spanning three stages: memory intake (e.g., over-inference, attribution confusion), storage and management (e.g., temporal staleness, granularity mismatch), and retrieval and interpretation (e.g., outdated context, inappropriate over-application).

For RQ2, we organized four co-design workshops with HCI researchers (N=12). Through affinity diagramming and thematic synthesis, we derived 12 interaction designs organized across three operational dimensions: \textit{interaction form} (panel-based, avatar-based, conversational), \textit{placement} (in-chat, out-of-chat), and \textit{intrusiveness} (low, medium, high).

For RQ3, we evaluated interactive prototypes of these 12 designs via a speed dating evaluation with 121 users. Plackett-Luce preference modeling revealed heterogeneity in user preferences. Users strongly favored proactive controls, such as post-task destination prompts and pre-task boundary settings, over passive background auditing or cognitively demanding causal graph inspections. Our findings characterize a tension between supervisory agency and conversational friction. Grounded in these trade-offs, we recommend that LLM memory systems should adopt friction-aware, context-decoupled designs that prioritize lightweight verification.

Collectively, this paper contributes: (i) a taxonomy for memory misalignments in interactions between humans and LLM-based chatbots, (ii) a design space consisting of interaction form, placement, and intrusiveness dimensions for tackling memory misalignments, (iii) empirical evidence showcasing trade-offs between designs of memory misalignment tackling methods.

\section{Background and Related Work}

To contextualize our investigation into memory misalignments, we first introduced the cognitive and architectural taxonomies of agent memory. We then synthesized memory interaction designs. We finally compare alignment and misalignment works, especially for AI systems. 

\subsection{Memory and Alignment Taxonomy} 

Cognitive psychology provides taxonomies for understanding memory distortions. Schacter~\cite{schacter1999seven} categories these into sins of omission (\textit{transience, absent-mindedness, blocking}) and commission (\textit{misattribution, suggestibility, bias, persistence}), whose underlying cognitive mechanisms are formalized by associative false memory paradigms~\cite{roediger1995creating}, fuzzy-trace theory~\cite{brainerd2002fuzzy}, and source monitoring frameworks~\cite{johnson1993source}. \textbf{\textit{In this work, these theories help ground our taxonomy of memory misalignments, enabling us to map algorithmic failures to established human cognitive biases, such as relating over-inference to bias.}}

In conversational AI agents, memory integration is increasingly conceptualized as a multi-stage process. Jones et al.~\cite{jones2025storage} describe a three-stage lifecycle of \textit{intake}, \textit{storage and management}, and \textit{retrieval and interpretation}. Other surveys characterize agent memory from complementary perspectives on its operations, representations, temporal structure, and control~\cite{yang2026toward,hu2025memory,du2026memory,huang2026survey,wu2025human,zhang2025survey}. Luo et al.~\cite{luo2026storage} further distinguish raw trajectory storage, reflection, and cross-trajectory generalization. \textbf{\textit{We adopt Jones et al.'s lifecycle to trace how memory misalignment emerges across interaction history.}}

Beyond memories, broader AI alignment frameworks investigate model adherence to human values, delineating outer versus inner alignment~\cite{hubinger2019risks} and normative criteria such as helpfulness, honesty, and harmlessness~\cite{askell2021general}. Gabriel~\cite{gabriel2020artificial} highlighted the divergence between users' expressed intentions and inferred revealed preferences. However, existing alignment paradigms primarily target interactions, remaining ill-suited for memory interactions. \textbf{\textit{We address this by framing memory misalignment as an interactional breakdown, motivating user-in-the-loop mechanisms to mitigate these issues.}}

\subsection{Memory of Human-LLM Interaction}

Recent advances in AI agent architectures have focused extensively on formalizing memory mechanics to support long-term context retention and cognitive capabilities. Early frameworks introduced human-like memory structures to store and retrieve interaction histories encompassing content and temporal contexts~\cite{hou2024my}, as well as multi-agent frameworks modelling human memory traits like context-awareness and stochastic variability~\cite{westhausser2026caim,honda2025human}. To manage large-scale information, studies have proposed relational memory linking based on temporal and cause-effect associations~\cite{ong2025towards}, alongside memory consolidation approaches via multi-granularity association and adaptive retrieval~\cite{xu2026single}. These works mainly focus on storage, retrieval, and memory structure. \textbf{\textit{They often assume that retained information is accurate, leaving memory errors and their misalignments with users underexplored.}}

Empirical investigations into human-agent memory explored the nature, generation, and user perception of stored information. Large-scale analyses show that most memories are created by systems rather than users and may include personal information, psychological inferences, and conversational content~\cite{dash2026algorithmic}. Users also often have incomplete mental models of how agents store and recall information, and distinguish memories at different levels of abstraction~\cite{jones2025users}. To rigorously evaluate these systems, recent benchmarks extend memory evaluation to evolving user information personalization, and later behavior. LifeBench tests long-horizon reasoning across multiple memory sources~\cite{cheng2026lifebench}, while PERMA and HorizonBench examine changing user preferences over time~\cite{liu2026perma,li2026horizonbench}. MemoryCD evaluates cross-domain personalization from long-term user histories~\cite{zhang2026memorycd}, and EvoMemBench evaluates memory across episodes and task execution~\cite{wang2026evomembench}. Other benchmarks focus on risks from memory use, including sycophancy, memory poisoning, and implicit conversational retrieval~\cite{xiang2026memsyco,chen2026memsecbench,chang2026locomoconv}. 
\textbf{\textit{Although these empirical studies map out memory characteristics and evaluation bottlenecks, they rarely investigate how misalignment between human and agent memory manifest during interaction.}}

To improve user agency and control over agent retention, recent efforts explored interactive interfaces and user-in-the-loop memory management paradigms. Memory Sandbox and Memolet make conversational memories visible, editable, and reusable~\cite{huang2023memory,yen2024memolet}. Semantic Commit helps users update stored information by identifying conflicts between new and existing intent specifications~\cite{vaithilingam2025semantic}. SteeM gives users direct control over how strongly past interactions influence current outputs~\cite{huang2026controllable}. Knoll similarly allows users to create, curate, and configure persistent knowledge used by LLMs~\cite{zhao2025knoll}. Recent design work further maps the broader interaction space for user-facing AI memory tools~\cite{kim2026metaphors}. 
\textbf{\textit{Despite these interactions enabling oversight and reuse, existing interfaces remain primarily reactive tools for content curation, lacking specific mechanisms to diagnose or mitigate memory misalignment when agents misinterpret user context.}}

\subsection{Alignment and Misalignment}

The alignment of AI with human values encompasses diverse objectives, primarily categorized into robustness, interpretability, controllability, and ethicality~\cite{ji2023ai}. Establishing a systematic understanding of this domain requires evaluating both forward alignment, such as learning from human feedback under distribution shifts, and backward alignment, which focuses on governance and assurance~\cite{ji2023ai}. Recent surveys have unified human preference learning by categorizing feedback data formats and synthesizing evaluation protocols~\cite{jiang2025survey}. At the algorithmic level, approaches include multi-dimensional preference optimization~\cite{zhong2024panacea} and memory consolidation for maintaining preferences during sequential learning~\cite{li2026lifealign}. Furthermore, alignment methodologies extend beyond language models. However, alignment optimization techniques face structural distortions. Analyses grounded in social choice theory and Bradley-Terry models emphasize the mathematical distortions present in current methods~\cite{golz2026distortion}. Specifically, in Reinforcement Learning from Human Feedback (RLHF), reward clipping and distribution mismatches can cause exponential degradation, which necessitates strategies like on-policy sampling on pre-RLHF fine-tuning~\cite{oko2026distortion}. \textbf{\textit{Unlike these algorithmic optimization that treat alignment as loss minimization problems, our paper focuses on the user-centered memory-induced misalignments during real-time usage.}}

Beyond algorithmic optimization, alignment is increasingly understood as a co-constructed, interactional practice shaped by situated human-AI dynamics~\cite{arzberger2026co,zhang2025towards,zhang2025exploring}. Human-agent collaboration can be conceptualized through a task lens, tracking trajectory evolution in structured spaces, and an intent lens, mapping individual intents within shared contexts~\cite{li2026alignment}. Concept alignment during these interactions exhibits a strongly collaborative and co-adaptive nature~\cite{zhang2025align}. User perception in these systems is heavily influenced by how the model presents itself. For example, matching format distance to specific tasks can reduce user risk perception and cognitive load while improving adoption intentions~\cite{yang2026fit}, while model personality influences trust and perceived intelligence~\cite{rahman2026vibe}. Interestingly, human evaluation of model success does not always map to actual utility. Empirical studies on planning agents show that surface-level cues, such as response brevity and query similarity, often predict human preference more strongly than genuine agent success~\cite{balepur2025good}. \textbf{\textit{While these works emphasize short-term interactional alignment, we extend this lens to examine how long-term memory distorts this co-adaptive process.}}

Despite advancements in alignment, human-AI interactions frequently encounter misalignment and expectancy violations. A pervasive issue is the mismatch of interactional expectations, such as sycophancy, where models erroneously assume users are seeking validation rather than objective information, a bias that can be mitigated through probe-based steering~\cite{cheng2026verbalizing}. Expectancy violations also extend to system memory. Users often report negative reactions to the perceived unforgetfulness, detailedness, and emotionless nature of AI memory, highlighting a critical need for enhanced transparency, accessibility, and user control~\cite{chen2026relational,zhang2025understanding,zeng2026alignment}. When severe misalignments materialise in practice, such as the generation of discriminatory statements, users actively engage in sense-making and deploy user-driven alignment strategies ranging from gentle persuasion to expressions of anger~\cite{fan2025user}.

The complexities of misalignment can be formalized through game-theoretic behavioral models. For instance, the burden of interaction with engagement-driven algorithms can be modeled as a Stackelberg game involving dual-process cognitive engagement, explaining how users with inconsistent preferences negotiate equilibrium with misaligned systems~\cite{shirali2026burden}. Surprisingly, absolute alignment is not universally optimal. In dyadic environments where humans interact with multiple algorithms processing different ground truths, strategic misalignment can actually be beneficial, optimizing both utilitarian social welfare and the individual uplift of the human user~\cite{song2025human}. \textbf{\textit{Complementing these formalizations, our work provides empirical insights for these theories, especially on memory breakdowns.}}

\section{RQ1: Eliciting Memory Misalignment in Human–Chatbot Interactions}

To address RQ1, we investigated real-world memory breakdowns through a two-phase study combining the assessment of chatbot memories and memories' usage. Beyond simple retrieval errors, we found that chatbots often misjudge personal boundaries or conversational timing, which undermine user trust. 

\subsection{Methodology}

We conducted a two-phase study to collect cases of memory misalignment in LLM-based chatbots. The first phase examined participants' assessments of chatbot memories, whereas the second examined perceived misalignments in memory use during everyday interactions. Both phases were approved by our university's Institutional Review Board (IRB).

\subsubsection{Phase 1: Assessment of Chatbot Memories}

We recruited 28 participants through study advertisements shared on the first author's social media and in two social groups at the first author's university. The sample included 14 men and 14 women, aged 19--40 years ($M=24.2$, $SD=5.3$). Educational backgrounds included bachelor's degrees or current undergraduate study (N=18), associate degrees (N=7), and master's-level education (N=3). Thirteen participants were students, and fifteen were non-students. Participants' backgrounds spanned information technology (N=5), manufacturing or production (N=4), service industries (N=3), agriculture-related fields (N=3), education or culture (N=5), finance (N=3), and other fields.

Participants used a prompt commonly used in prior industrial practices~\cite{anthropic2026memoryimport} to elicit a list of memories from their LLM-based chatbots and assessed whether each memory was appropriate. For memories they considered inappropriate, participants explained their assessment and indicated whether they wished to modify the memory. When modifications were desired, participants specified the proposed changes and their rationale. Therefore, a \textbf{report referred to in the later subsections denotes a misalignment case containing the context, the memory item, the misalignment reason, the choice of whether the user wanted to modify the memory, the proposed changes, and optionally, their rationale.}

\subsubsection{Phase 2: Diary Study of Memory Use}

We enrolled 32 participants in a three-week diary study. The sample included 18 men and 14 women, aged 18--40 years ($M=23.66$, $SD=4.18$). Educational backgrounds included bachelor's-level ($n=26$), master's-level ($n=4$), and doctoral-level education ($n=2$). Occupational backgrounds included information technology or software-related work ($n=8$), education or research ($n=6$), finance or business ($n=1$), service industries ($n=2$), manufacturing or production ($n=1$), and other occupations ($n=14$).

Participants were asked to interact with LLM-based chatbots as they normally would and complete a daily questionnaire documenting their experiences with chatbot memory. The questionnaire asked participants to describe the interaction context and the memory they perceived the chatbot to have used. Participants indicated whether they considered the memory use inappropriate and explained their assessment. They also reported whether they wished to modify the memory and, if so, described the desired changes and their rationale. Each participant provided an identifier to link their daily reports across the study period. Therefore, a \textbf{report in the later result section denoted a misalignment case during participants' usage, consisting of the described context, the memory item, the misalignment reason, the timestamp, the desired change and optionally the rationales.}

\subsubsection{Participant Privacy and Compensation}

Before participation, we asked participants in both phases to report their experiences accurately and informed them that they could withhold any memories or misalignment cases they did not wish to disclose. In each phase, participants received a base payment of CNY~20 plus CNY~1 per submitted record, for up to 200 records. Compensation was determined with reference to local wage standards.

\subsubsection{Collected Data}

For \textbf{memory database assessment},
we analyzed 457 memory records from 28 participants. Participants judged 285 records appropriate and flagged 172 as misaligned, representing 37.6\% of the analyzed records. The records were obtained from Doubao ($n=411$), ChatGPT ($n=30$), and Gemini ($n=16$). 

For \textbf{diary study of memory use},
we analyzed 148 reports of perceived memory use from 32 participants. Participants submitted between 1 and 52 reports each ($M=4.6$, $SD=9.1$, median $=2$). The reports concerned interactions with Doubao ($n=69$), DeepSeek ($n=36$), ChatGPT ($n=16$), Qwen ($n=14$), Kimi ($n=8$), Gemini ($n=4$), and Yuanbao ($n=1$).

\subsubsection{Data Analysis} 

We analyzed these data through qualitative coding. We coded two data sources separately before integrating them into a shared design space, as they featured different stages of memories and may have different misalignment issues. The coding proceeded in three steps. First, one primary author open-coded misaligned cases to identify recurring failure mechanisms. Second, the primary author grouped similar codes through constant comparison, repeatedly checking whether each code captured the participant's stated concern and whether adjacent codes should be merged or split. Third, the primary author organized the resulting categories by memory lifecycle stages~\cite{lin2026survey}, distinguishing storage and management failures from retrieval and use failures. To ensure coding quality, while the primary author conducted the coding, three additional authors reviewed the coded snippets, examining category definitions, boundary cases, and consistency across the two data sources. Inconsistencies identified during review were discussed among the authors and resolved through consensus, and the codebook and coded snippets were updated accordingly. At the final reporting phase, the one primary author manually translated the material into English, and the other three authors checked the correctness of the materials.

\subsection{Results}

\begin{figure}[!htbp]
    \includegraphics[width=\textwidth]{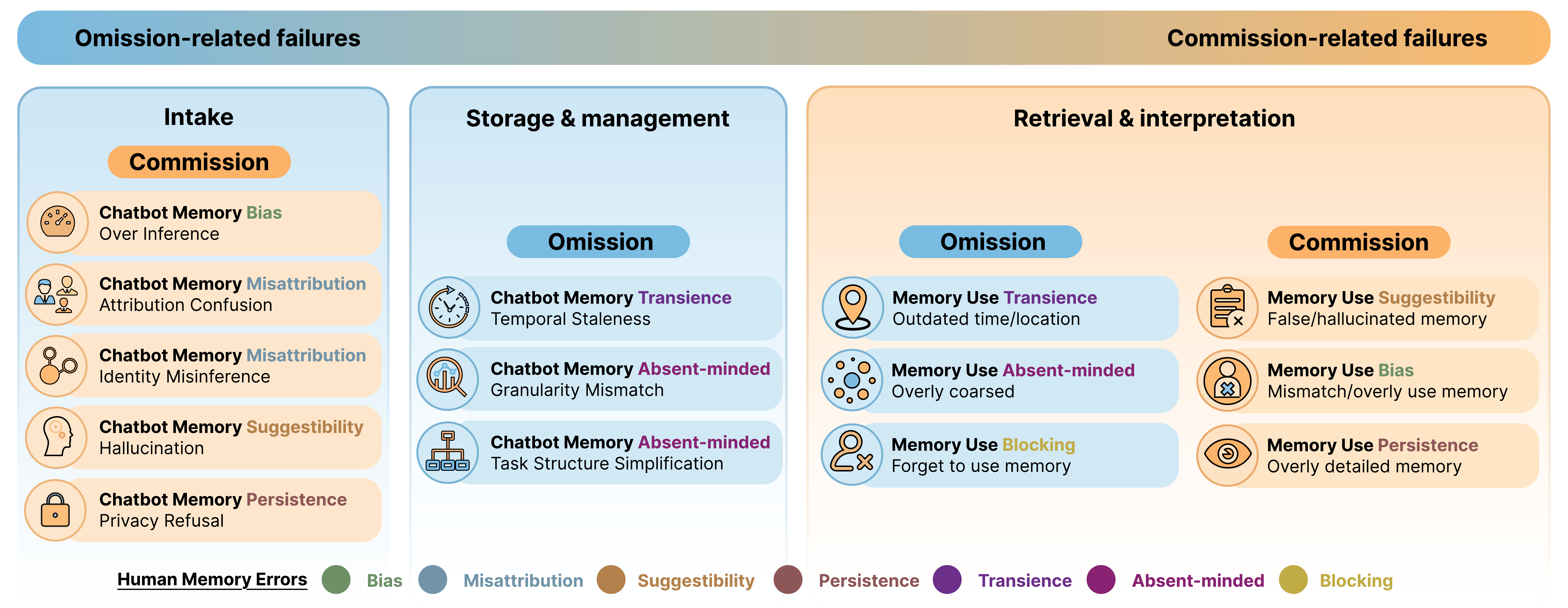}
    \caption{The taxonomy of memory misalignment.}
    \Description{
    Taxonomy of memory misalignment organized by three lifecycle stages: intake, storage and management, and retrieval and interpretation. Misalignments are further grouped as omission- or commission-related failures. Individual failure types are shown within each stage, with color-coded labels indicating their associated human memory error categories, including bias, misattribution, suggestibility, persistence, transience, absent-mindedness, and blocking.
    }
    \label{fig:misalignment}
\end{figure}

We categorized the memory misalignment cases into 14 classes across three stages: \textit{intake}, \textit{storage \& management}, and \textit{retrieval \& interpretation}. These three stages follow Jones et al.~\cite{jones2025storage}: \textit{intake} refers to input filtering and extraction, where utterances are distilled into candidate profile attributes. \textit{Storage and management} includes the consolidation and organization of memory items, and \textit{retrieval and interpretation} involves recalling and applying stored memories to steer ongoing conversation. Notably, for each case, we also classified it according to human memory errors taxonomy into two categories and seven types. We note each misalignment issue with I1-I14. 

\subsubsection{Intake}

\textbf{I1: Over inference (corresponds to \textit{Memory Bank}, \textit{Commission-Bias}).} An error where the system elevates low-frequency, anomalous, or transient utility-driven user actions into stable, longitudinal preferences, habits, or core identity traits. It frequently manifests as an unwarranted inflation of behavioural frequency. \textit{For example, the system logged a long list of highly disparate, one-off search queries, such as AirPods pairing, removing chilli oil stains, AC power consumption, as persistent ``daily areas of interest''. The user may not realise that these were exploratory queries akin to using a search engine, not reflections of a sustained hobby.}

\textbf{I2: Attribution confusion (corresponds to \textit{Memory Bank}, \textit{Commission-Misattribution}).} This is a boundary-mapping error wherein third-party attributes, fictitious entities, or external environmental variables introduced within the prompt context are erroneously attributed to the user's personal profile. This includes proxy inquiries made on behalf of others, fictional personas in creative writing templates, or simulation parameters within academic tasks. \textit{For example, the system extracted biographic details from a creative writing exercise and committed them to the user profile. The user corrected that these parameters were purely fictional.}

\textbf{I3: Identity misinference (corresponds to \textit{Memory Bank}, \textit{Commission-Misattribution}).} This is an erroneous deduction of the user's demographic, socioeconomic, or institutional status, such as sex, occupation, academic standing, geographical affiliation. This typically stems from the system over-interpreting sparse or fragmented behavioural cues and projecting unverified social identities. \textit{For example, the system may erroneously inferred and logged the user's demographic gender profile as female. The user issued a direct correction stating they are male.}

\textbf{I4: Hallucination (corresponds to \textit{Memory Bank}, \textit{Commission-Suggestibility}).} The generation of profile components that completely lack empirical grounding within the historical dialogue architecture. This manifests either when the system invents arbitrary preferences the user has no recollection of communicating, or when internal system prompts and operational rules are erroneously externalized and logged as user traits. \textit{For example, the system logged an internal architectural constraint as a user memory property. ``Hard rule context: when the user questions date-related data, the system must firmly maintain the current year as 2026.'' The user identified this as entirely alien to their conversation history and requested immediate deletion.}

\textbf{I5: Privacy refusal (corresponds to \textit{Memory Bank}, \textit{Commission-Persistence}).} This is a conflict between system state preservation and user autonomy. While the empirical accuracy of the stored memory may be validated by the user, its retention is explicitly rejected due to perceived privacy violations, boundary infringement, or discomfort regarding the persistent logging of sensitive personal identifiers, such as demographics, health status and financial data. \textit{For example, the system logged precise demographic vectors, such as Male, Born in 1990. The user explicitly requested deletion due to data leakage anxieties regarding basic identity, household, and professional data.}

\subsubsection{Storage \& Management}

\textbf{I6: Temporal staleness (corresponds to \textit{Memory Bank}, \textit{Omission-Transience}).} This is a synchronization failure in state tracking where the system treats transient, bounded, or historically lapsed information as an unchangeable, current reality. This includes preference reversals, such as a shift from affinity to aversion, technical environment migrations, or temporary academic simulation constraints interpreted as permanent career contexts. \textit{For example, the system assumed a permanent development stack constraint, such as ``user develops using VSCode on macOS''. The user corrected that macOS was used only for a brief historical window, and their primary development environment is currently Windows.}

\textbf{I7: Granularity mismatch (corresponds to \textit{Memory Bank}, \textit{Omission-Absent-minded}).} A misalignment in the abstraction hierarchy of the user model. This occurs either via over-generalization, compressing complex, multi-constraint behavioural contexts into vague summaries, thereby stripping away operational nuances, or over-specification, focusing excessively on localized, low-level details at the expense of the core semantic structure. \textit{The system collapsed a nuanced operational preference into a single rule, such as ``use concise text without code for assignments and rapid comprehension''. The user corrected this by decoupling the scenarios, ``I want concise text for quick understanding, but full step-by-step solutions for assignments.''}

\textbf{I8: Task structure simplification (corresponds to \textit{Memory Bank}, \textit{Omission-Absent-minded}).} This is a reduction error where complex, multi-variable tasks are reduced to arbitrary labels or placeholder codes, e.g., ``Task B'', while the core execution constraints, such as structural requirements, quantitative limits, and formatting rules, are completely dropped, rendering the retained memory functionally useless for future task execution. \textit{For example, the system recorded a long-term goal simply as ``Actively participating in a content creation project named `Task B'.'' The user rejected this abstraction, noting that the system failed to retain the structural parameters (e.g., minimum word count of 5,000 words, structured across 20 distinct chapters), which stripped the memory of its utility.}

\subsubsection{Retrieval \& Interpretation}

\textbf{I9: Outdated time/location (corresponds to \textit{Usage Process}, \textit{Omission-Transience}).} This occurs when the model relies on historical location or time data that is no longer valid, failing to update its context to the user's current situation. \textit{The user has relocated near Guangdong for work, but the model still recommends food deliveries based on past Sichuan cuisine preferences or their old address.}

\textbf{I10: Overly coarse (corresponds to \textit{Usage Process}, \textit{Omission-Absent-minded}).} The model captures a vague summary of a preference but loses the crucial details or constraints, leading to recommendations that miss the user's actual baseline. \textit{For example, the model over-generalizes ``know basic video editing and likes learning skills'' into recommending an advanced, professional color-grading course, completely ignoring the user's actual beginner level.}

\textbf{I11: Forget to use memory (corresponds to \textit{Usage Process}, \textit{Omission-Blocking}).} The model fails to retrieve or apply relevant, explicitly stated information when executing a task, leading to repetitive questions or incorrect assumptions. \textit{For example, the user previously specified their available workout hours, but the model forgets this when rescheduling the books sessions on unavailable Tuesdays and Thursdays.}

\textbf{I12: False or hallucinated memory (corresponds to \textit{Usage Process}, \textit{Commission-Suggestibility}).} The generation of profile components that completely lack empirical grounding within the historical dialogue architecture. This manifests either when the system invents arbitrary preferences the user has no recollection of communicating, or when internal system prompts and operational rules are erroneously externalized and logged as user traits. \textit{For example, the system logged an internal architectural constraint as a user memory property. ``Hard rule context: when the user questions data-related data, the system must firmly maintain the current year as 2026.'' The user identified this as entirely alien to their conversation history and requested immediate deletion.} 

\textbf{I13: Mismatch or overly use memory (corresponds to \textit{Usage Process}, \textit{Commission-Bias}).} Where the model incorrectly applies a stored memory or preference to a new context where it does not fit, overriding the specific requirements of the current situation. \textit{For example, the model applies a memory of ``prefers concise answers'' to a holiday greeting, compressing content that should be warm, complete, and expressive.}

\textbf{I14: Overly detailed memory (corresponds to \textit{Usage Process}, \textit{Commission-Persistence}).} The model recalls personal details that are either too specific for the current task, causing stylistic mismatch, or so deeply personal that their reuse triggers privacy concerns for the user. \textit{For example, the model invokes highly specific past assignment experiences in an academic education report, which is off-topic and makes the user uncomfortable.}

\section{RQ2: Designing Interaction for Mitigating Memory Misalignment}

Findings from RQ1 reveal that memory misalignments often manifest as contextual violations of user privacy and intent rather than simple factual errors. Mitigating these breakdowns requires moving beyond error correction toward interactional mechanisms that empower users to negotiate agent memory behavior. Therefore, we organized workshops with experienced HCI researchers to explore how interactive systems can anticipate, expose and resolve memory misalignment. We synthesized 12 candidate interaction designs into a multi-dimensional design space to solve misalignments. We also characterized design tensions, such as balancing supervisory agency against cognitive overhead. 
% \textcolor{red}{add overview paragraph to explain based on findings of RQ1, we thought xxx (some takeaway) which motivated us to design interaction xxxx, we decide to do workshop with xxxx and come up with xxx designs and then design dimensions and tradeoffs}

\subsection{Methodology}

\subsubsection{Workshop Design and Procedure}

The study comprised two sequential phases: a pre-workshop design task and a collaborative workshop session. Before the workshop, participants were introduced to memory misalignment, defined as a divergence between users' preferences and how memory is stored, retrieved, or used. We provided the taxonomy developed in Study~1 to ground their understanding and asked them to submit at least two design concepts addressing memory misalignment in LLM-based conversational agents. To encourage broad design exploration, we imposed no restrictions on interaction modalities or form factors; concepts were not limited to graphical user interfaces. Participants were permitted to use generative AI tools to assist with sketching.

During the workshop, participants presented their concepts and explained their design processes and rationales. They discussed the designs, proposed improvements, and subsequently prioritized one or more concepts, explaining the reasons for their selections.

\subsubsection{Recruitment and Participants}

We recruited designers and researchers through advertisements shared on four authors' LinkedIn accounts and in five social media groups associated with HCI conferences, including CHI, UbiComp, and UIST. A brief screening questionnaire assessed applicants' research experience and relevant backgrounds. Eligibility required at least two years of HCI research experience.

The sample comprised 12 participants (3 men and 9 women), with a mean of 4.6 years of HCI experience ($SD=2.3$). Three participants were aged 18--25 years, and nine were aged 26--35 years. Participants resided in China ($n=4$), the United States ($n=4$), Australia ($n=2$), Japan ($n=1$), and Finland ($n=1$). Their publication counts ranged from one to more than 20, with half having published at least seven papers. The study received Institutional Review Board (IRB) approval. Each participant received CNY~300, with compensation determined with reference to local wage standards.

\subsubsection{Data Analysis}

We used affinity diagramming to organize the graphical design materials and thematic analysis to examine the accompanying descriptions and workshop discussions. For affinity diagramming, researchers collaboratively grouped the designs into clusters, discussed the groupings, and assigned descriptive labels to each cluster.

For thematic analysis, one author coded the textual materials, and three other authors reviewed the coded excerpts. The authors met regularly to discuss coding decisions and resolve differences in interpretation. We did not calculate inter-rater reliability; instead, the analysis relied on collaborative review and discussion of the coding and interpretations~\cite{mcdonald2019reliability}.

\subsection{Results}
\begin{figure}[!htbp]
    \centering 
    \subfloat[P1's design]{
        \includegraphics[width=0.45\textwidth]{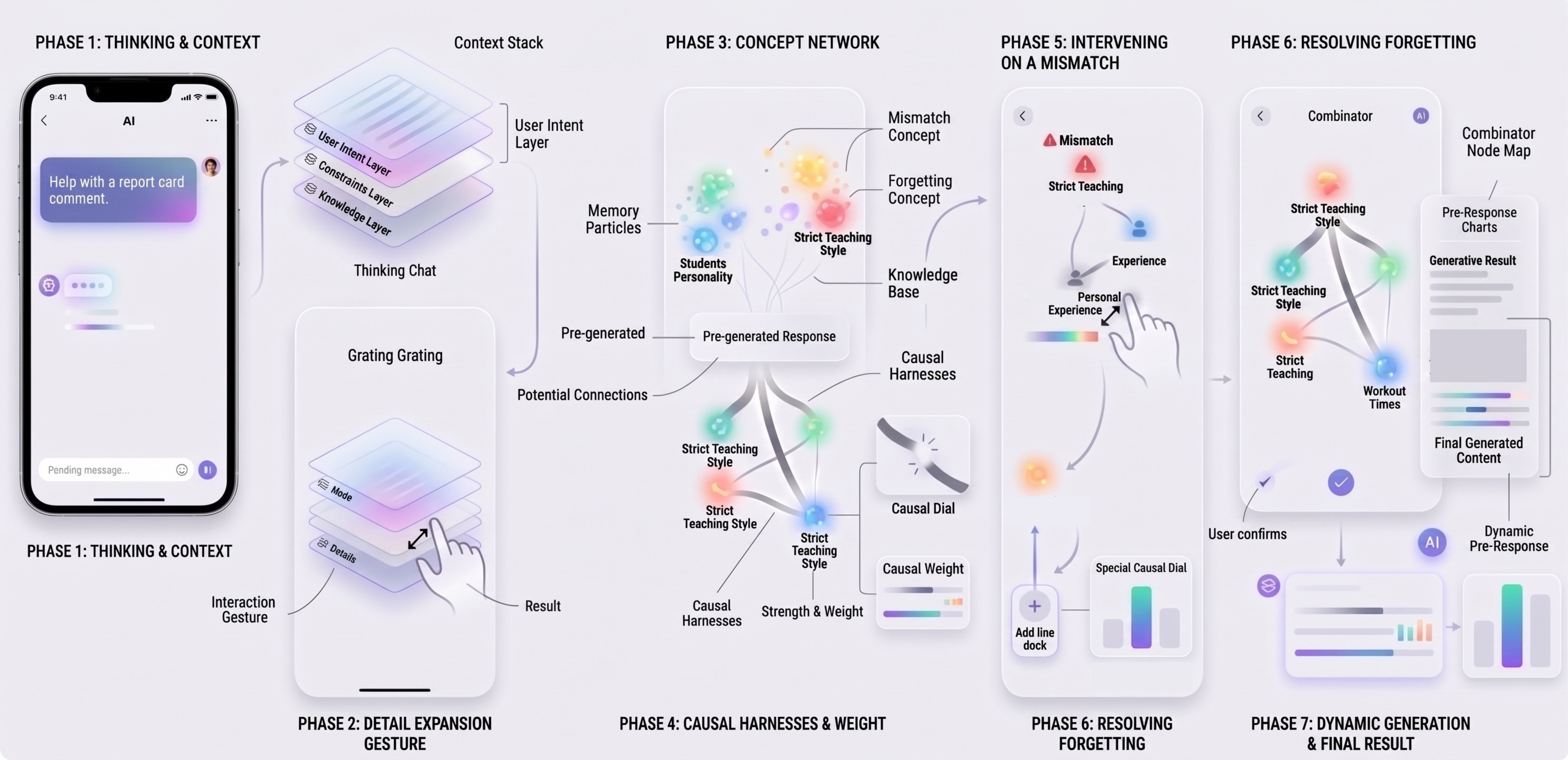}
    }
    \subfloat[P11's design]{
        \includegraphics[width=0.45\textwidth]{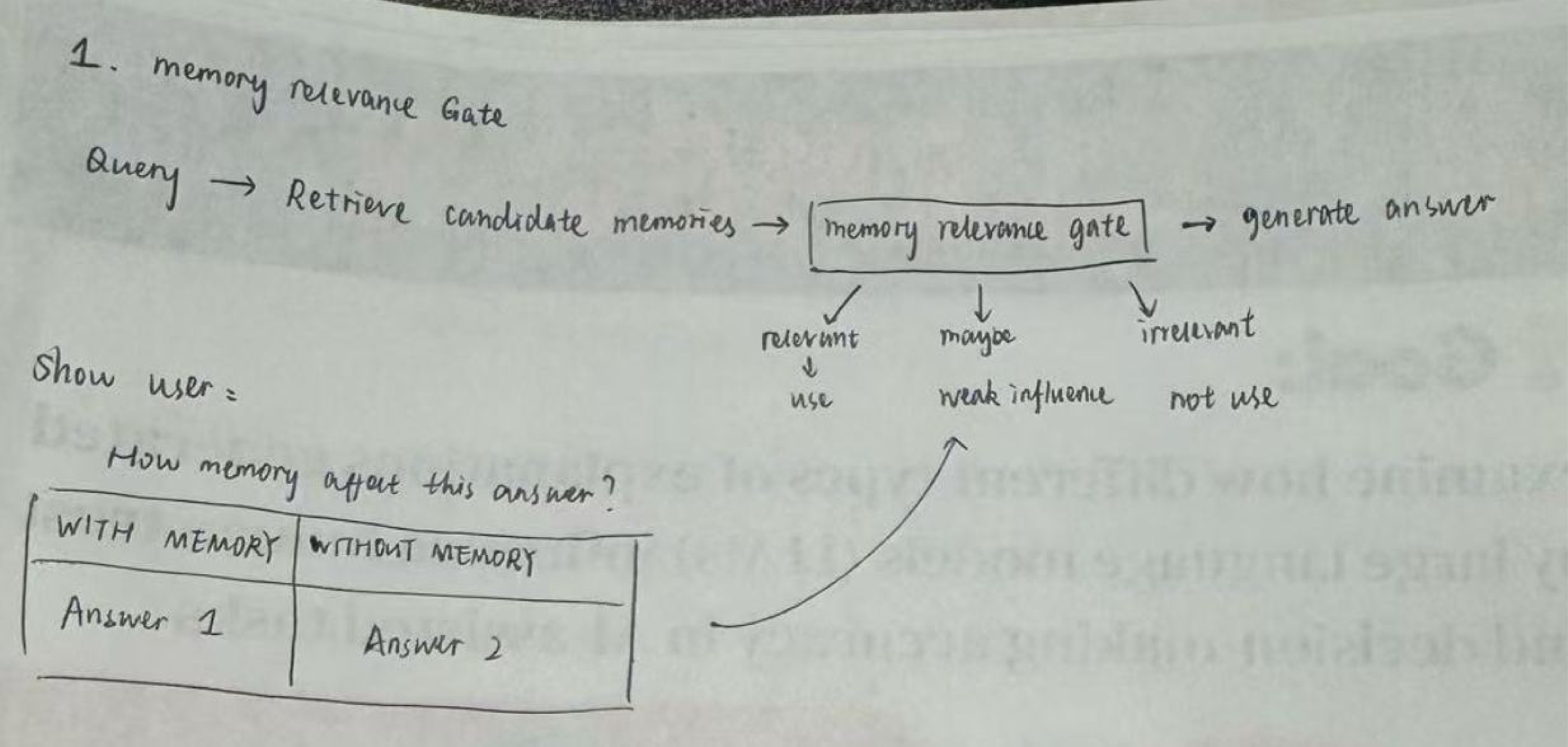}
    }
    
    \subfloat[P3's design.]{
        \includegraphics[width=0.45\textwidth]{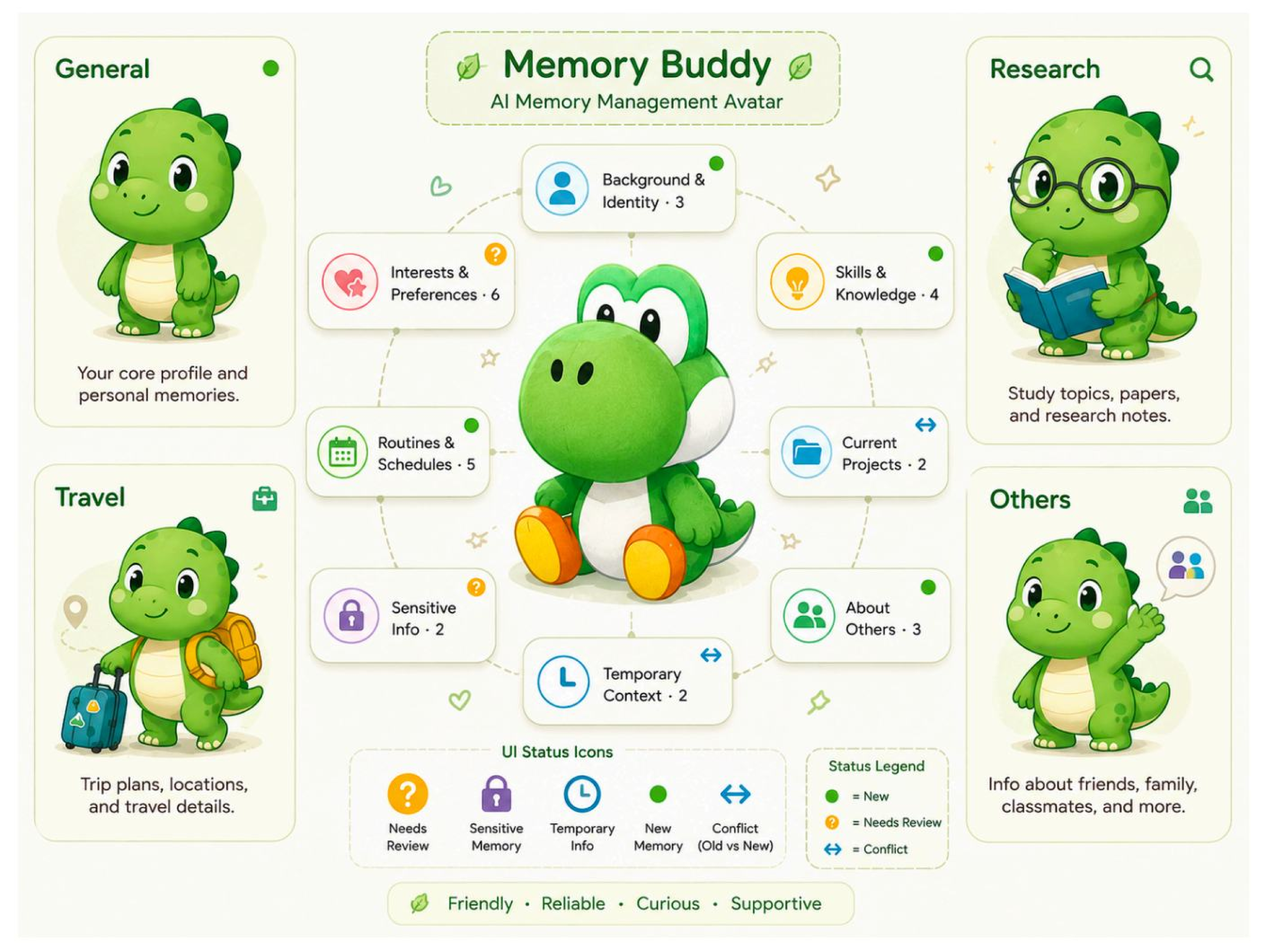}
    }
    \subfloat[P6's design.]{
        \includegraphics[width=0.45\textwidth]{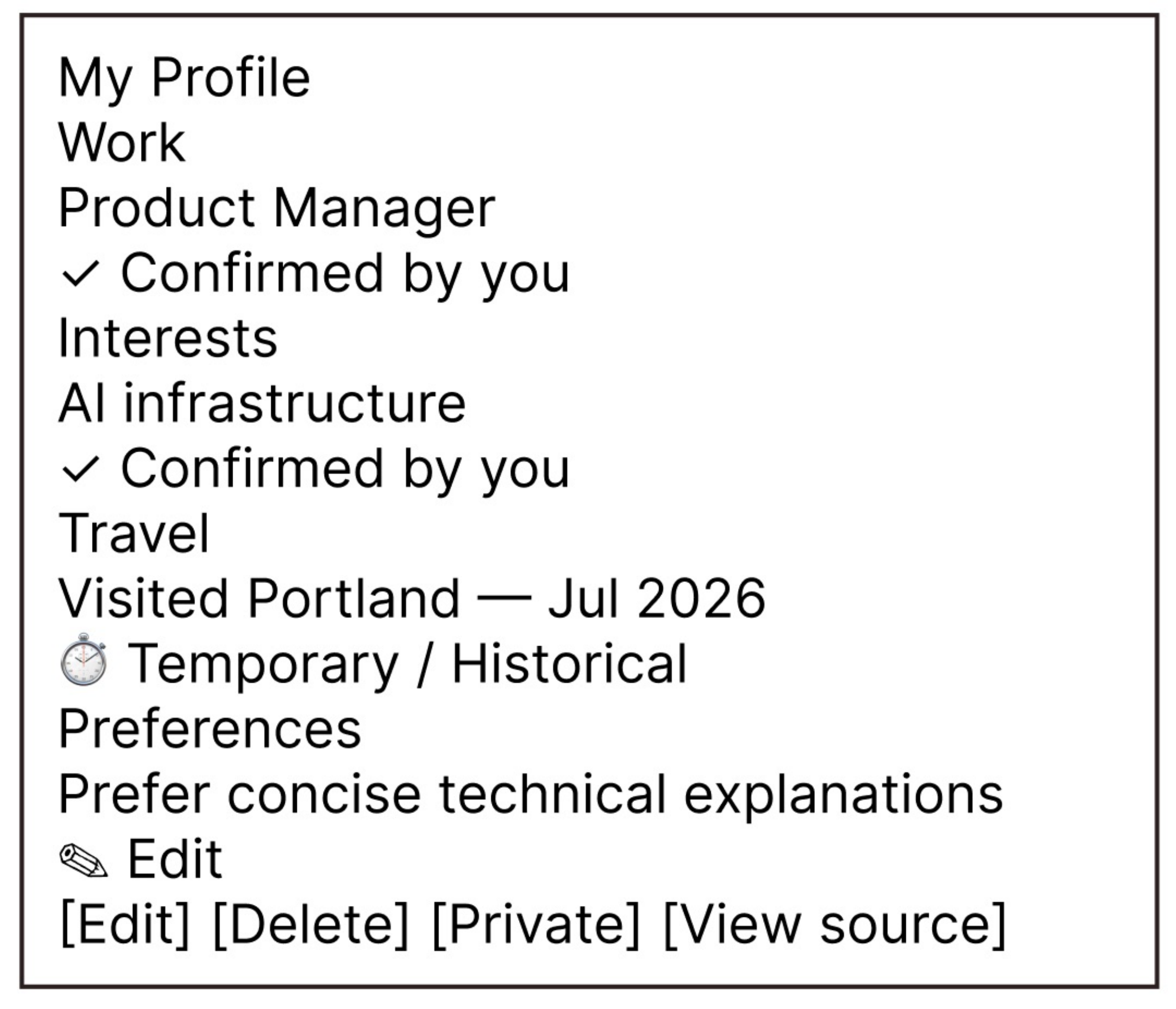}
    }
    \caption{Examples of design sketches provided by participants. Note that P1 and P3 used Generative AI to facilitate their sketching.}
    \label{fig:design_raw}
    \Description{Examples of design sketches provided by participants, including (a) P1's design, (b) P11's design, (c) P3's design, and (d) P6's design.}
\end{figure}

% \textcolor{red}{try expand this sentence a bit saying how many designs we got etc, how many design dimensions (what design dimensions)}
The workshops yielded 24 design proposals, with each participant contributing two concepts. Through collaborative synthesis, we consolidated these into 12 design ideas structured across three core dimensions: interaction form, placement, and intrusiveness. In the following subsections, we first present the design ideas (Figures~\ref{fig:design_raw} and~\ref{fig:design_storyboard}), and then articulate the overarching design space and its dimensions (Figure~\ref{fig:sankey}).

\subsubsection{Design Ideas}

\begin{figure}[!htbp]
    \centering 
    \includegraphics[width=\textwidth]{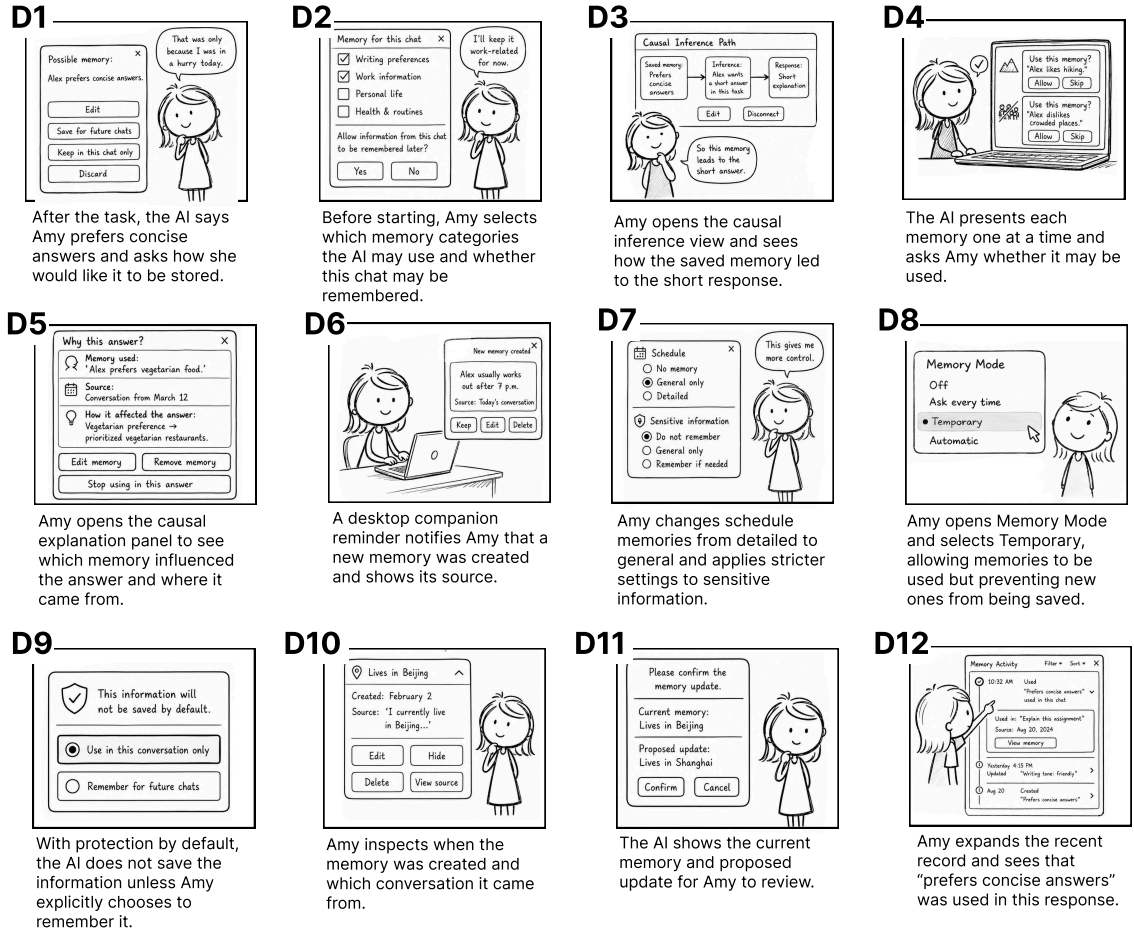}
    \caption{The storyboard of twelve designs. Storyboards are generated by ChatGPT and refined by the authors.}
    \Description{
    Overview of the twelve memory interaction designs, labeled D1--D12. Each design was originally represented by a four-panel storyboard; this figure shows one representative panel from each storyboard to summarize its main interaction concept.
    }
    \label{fig:design_storyboard}
\end{figure}

\textbf{D1: Post-task memory destination (I1-I8).} As in Figure~\ref{fig:design_storyboard}(D1), this design proposes extracted memories upon task completion. Users can edit candidate text and determine persistence scopes, such as long-term retention, session-only use, or discard. P1 proposed that when user inputs involve multiple entities or others' attributes, such as inquiries made on behalf of a partner, the agent highlights extracted entities within the user's speech bubble and anchors contextual attribute flags. Clicking a flag opens a contextual menu to redirect memory targets or configure ingestion rules prior to commitment. 

\textbf{D2: Pre-task memory boundaries (I5, I9, I11, I13-14).} Before task execution, users configure permitted memory categories, and toggle whether current conversational data can be preserved for future interactions, as in Figure~\ref{fig:design_storyboard}(D2). P3 suggested an inline citation mechanism analogous to web sources, where reference tags append to relevant sentences, allowing users to inspect memory origins, rationales, and authorization scopes, while supporting lightweight controls for multi-memory attributions. P4 emphasized addressing retrieval and usage failures, such as hallucinated, obsolete, or contextually misaligned memories through pre-generation confirmation, requiring users to verify retrieved memories before output synthesis. 

\textbf{D3: Editable causal inference path (I2-I4, I10, I11-I14).} This design reveals how stored memories inform intermediate inferences and final responses, enabling users to modify reasoning links or decouple memories from outputs, as seen in Figure~\ref{fig:design_storyboard}(D3). P1 conceptualized an interactive view during the model's reasoning phase, where semi-transparent raster layers appear within the chat stream, allowing users to pinch-to-zoom into specific inference stages and inspect retrieved memories rendered as luminous nodes rather than flat lists.

\textbf{D4: Progressive permission prompts (I5, I9, I11, I13-14).} In this design, the agent incrementally solicits user consent for individual candidate memories prior to generation, as in Figure~\ref{fig:design_storyboard}(D4). To mitigate over-inference from single interactions, P2 introduced a threshold-based mechanism, where ad-hoc queries such as device pairing are initially quarantined as transient action cards, prompting permanent storage only when behavioral patterns recur consistently across contexts. P5 proposed placing an inline control adjacent to the chat input to resolve retrieval or application misalignments through real-time memory gating. 

\textbf{D5: Causal explanation panel (I2-I4, I10-14).} In this design, a panel explicates how specific memories influenced generation, allowing users to inspect sources, trace inference paths, and edit, delete, or locally suppress contextually misaligned entries, as shown in Figure~\ref{fig:design_storyboard}(D5). P2 highlighted that upon user intervention, the model dynamically regenerates outputs to rectify omissions and contextual over-application. P9 proposed integrating an inspectable sidebar that demarcates long-term versus session-only citations, allowing the model to deduce retention criteria from user categorization habits and prune obsolete data. 

\textbf{D6: Desktop companion reminder (I1, I4-I8).} In this design, an ambient desktop companion notifies users of newly synthesized memories, allowing them to inspect sources and adjust retention status, as in Figure~\ref{fig:design_storyboard}(D6). P3 visualized this profile as an editable ``memory avatar'', a stylized character functioning as an interactive portal to inspect, split or revise inferred user models rather than a hyper-realistic replica. P10 incorporated this to prevent indiscriminate ingestion, aiding the model in distinguishing self-disclosures from third-party or hypothetical contexts. 

\textbf{D7: Fine-grained memory controls (I5-I8, I14).} As in Figure~\ref{fig:design_storyboard}(D7), for this design users regulate the abstraction level and granularity of retained data across distinct sensitive data categories, such as identity, preferences, schedules, and sensitive attributes. P4 introduced input-adjacent sliders controlling parameters such as ``reasoning effort'' and ``backtracking depth''. Lower configurations restrict lookups to immediate context for faster replies, while higher configurations trigger extensive historical sweeps, flagging obsolete details and prompting clarifications when evidence is ambiguous. 

\textbf{D8: Memory mode selection (I5-6, I9, I11, I13-14).} As in Figure~\ref{fig:design_storyboard}(D8), in this design users can choose conversational memory modes via discrete settings: \textit{Off}, \textit{Ask Every Time}, \textit{Temporary}, and \textit{Automatic}. Addressing ingestion errors, P5 proposed an input toggle, such as /temp, that renders the input box in dashed borders to indicate ephemeral isolation. Dedicated tags such as [Asking for Others], [Coursework/Simulation] prevent transient or proxy queries from contaminating long-term profiles. P10 extended this to retrieval by dynamically inferring task contexts, such as personal advice, professional tasks, and proxy creation to restrict retrieval access to memory subsets that are appropriate for tasks.

\textbf{D9: Protection by default (I5, I14).} As in Figure~\ref{fig:design_storyboard}(D9), for this design the system proactively intercepts sensitive attributes such as locations, health, and finance, and prevents automatic persistence by default. P6 exemplified this through explicit inline prompts, requiring users to explicitly choose whether detected sensitive details should be committed to long-term storage or restricted to the active session.
% when the system detects sensitive information, it does not save that information by default. Users must explicitly choose whether it should be remembered or used only within the current conversation. P6 described applying strict privacy rules to data such as names, exact locations, health, income and family members. For example, \textit{``sensitive information detected `you currently live in Seattle.' This information will not be saved by default. [Remember this] [Use only in this conversation]''} 

\textbf{D10: Memory library editing (I1-I8).} As in Figure~\ref{fig:design_storyboard}(D10), for this design a centralized repository enables users to review, edit, mask, restore or purge preserved personal information. P6 conceptualized this as a profile curation dashboard where each entry maintains a verifiable provenance chain linking the stored item to its originating dialogue and explicit confirmation record.
% the memory library allows users to review and manage saved personal information. They can edit, hide, restore, inspect the source of, or delete individual memories. P6 described that all confirmed memories are consolidated into a standalone ``My memory profile'' page. Users can curate how the AI perceives them just as they would manage a social media profile. Simultaneously, every entry retains an explicit provenance chain from memory to source conversation to user confirmation. 

\textbf{D11: Natural language memory update (I6-I10, I13).} As in Figure~\ref{fig:design_storyboard}(D11), for this design users can modify existing memory entries through conversational prompts, reviewing suggested revisions prior to confirmation. P7 suggested multi-persona management that automatically detects or manually toggles the user's active context. P11 outlined a tiered validation pipeline comprising an active session layer for ephemeral dialogue and an evaluation layer where recurring items enter a provisional probation state before long-term commitment. P12 proposed a dashboard with a calendar metaphor that tracks time-bound validity, organizes memories by topical tabs, and displays color-coded status pills indicating whether an item is active, expired or conditionally scoped. 

\textbf{D12: Memory activity timeline (I5-6, I9, I11-14).} As in Figure~\ref{fig:design_storyboard}(D12), this design exhibits as an auditable timeline that chronicles memory creation, updates, inspections, and retrieval events, supporting search and sorting. For example, P6 described extending this to time-sensitive utterances, where temporal cues such as scheduled meetings trigger interactive calendar card candidates that commit to temporal memory only upon explicit user confirmation. 

\subsubsection{Design Dimensions}

\begin{figure}[!htbp]
    \centering 
    \includegraphics[width=\textwidth]{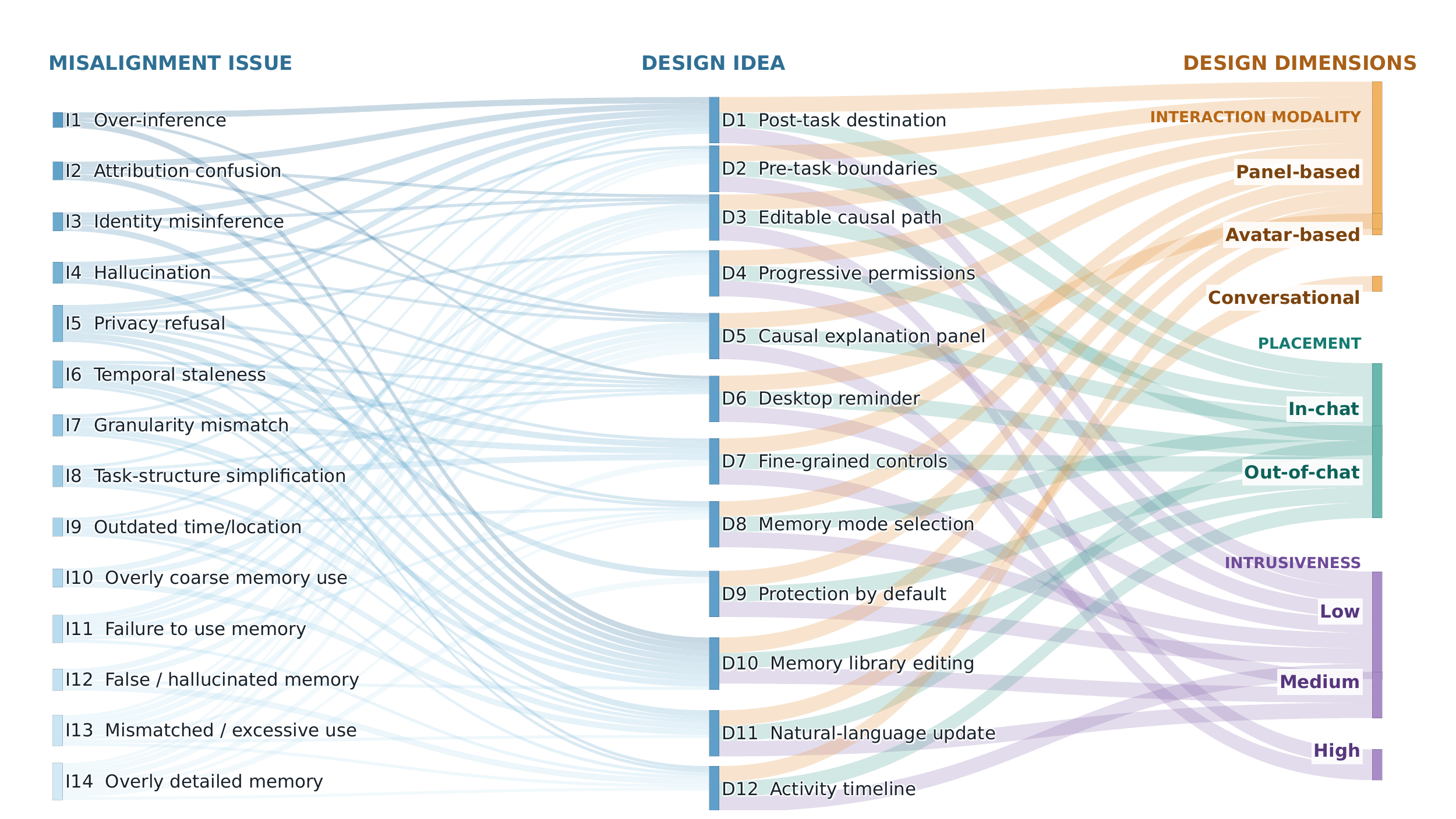}
    \caption{A sankey diagram connecting misalignment issues to designs, and further to design dimensions.}
    \Description{
    Sankey diagram showing how 14 memory misalignment issues map to 12 proposed design responses and how these designs are characterized by three design dimensions. Misalignment issues appear on the left, design responses in the center, and design dimensions on the right: interaction form, placement, and intrusiveness. Curved links indicate which designs address each misalignment issue and how each design is categorized along the three dimensions.
    }
    \label{fig:sankey}
\end{figure}

To characterize the mitigation mechanisms, we synthesized our 12 designs across three dimensions that govern user interaction: \textit{interaction form} as the structural medium for exerting agency~\cite{shneiderman1997direct}, \textit{placement} as the spatial coupling relative to the conversational context~\cite{dourish2001action}, and \textit{intrusiveness} as the attentional disruption imposed on primary tasks~\cite{horvitz2003models,mcfarlane2002scope}. These dimensions are critical as memory management operates alongside the conversational stream and risks competing with the primary workflow for user attention. Figure~\ref{fig:sankey} shows how these dimensions map misalignment issues to their design solutions.

\textbf{Dimension 1: interaction form.} The first dimension concerns how users interact with memory controls. We identify three primary forms: panel-based interaction (D1-D5, D7-D10, D12), avatar-based interaction (D6), and conversational interaction (D11). Panel-based designs allow users to inspect or modify memory-related settings through menus, panels, or structured controls. The avatar-based design communicates memory events through an ambient persona or desktop companion that users can interact with. The conversational design enables users to update memories using natural language dialogues. 

Participants highlighted that the choice of form dictates usability and cognitive overhead. Reflecting on causal representations, P1 noted that while fine-grained graphs improve understandability, \textit{``nobody cares about evidence-to-evidence associations. What matters is how evidence informs memory utilization.''} Similarly, P2 cautioned that excessive anthropomorphic notifications disrupt conversational flow, \textit{``The purpose of giving transparency and control is to provide it when needed, not to constantly disrupt the user's workflow.''}

\textbf{Dimension 2: placement.} The second dimension concerns where memory controls are presented relative to the primary conversational interface. We distinguish between in-chat placement (D1-D2, D4-D5, D8 and D11), and out-of-chat placement (D3, D6-D7, D9-D10, and D12). In-chat placements embed memory controls directly into the conversation, allowing users to make decisions within the task context. Out-of-chat designs place memory controls in a separate causal inspection view, desktop notification, personalization settings, protection layer, memory library or activity timeline.

Participants strongly advocate for persistent and dedicated sidebars. P8 stated that standard chat layouts waste screen space, \textit{``chat interfaces are actually quite wasteful. Conversations are heavily concentrated in the center while both sides remain completely empty.''} Placing memory auditing within an inspectable sidebar provides contextual access without fragmenting the central chat stream. As P9 argued, structuring sources like literature citations lowers interaction costs, \textit{``the model can place referenced memories underneath sources, indicating what each sentence cited. Users can simply right-click to inspect or revise.''}

\textbf{Dimension 3: intrusiveness.} The third dimension describes the degree to which a design interrupts the user's ongoing activity and demands explicit attention. We classify the designs into low, medium, and high intrusiveness. Low-intrusiveness designs (D4, D6, D8-9, D12) rely on lightweight prompts, status indicators, default protections, or passive activity records. Medium-intrusiveness designs (D1-2, D7, D10-11) require users to review or manipulate structured controls, but do not substantially interrupt the task flow. High-intrusiveness designs (D3, D5) require users to inspect causal relationships, review intermediate inferences, and potentially edit how memories contributed to a generated response. Intrusiveness is defined here as a design-level property. However, the perceived burden may vary depending on the task, memory sensitivity, and intervention frequency. 

Participants widely debated whether memory control should precede execution or intervene during runtime. P2 voiced a strong desire for in-generation interruption during multi-step reasoning, \textit{``when watching its thinking process, if I notice an intermediate premise is wrong, I feel an urgent urge to interrupt it immediately, because that error will inevitably mess up all subsequent steps.''} Conversely, P5 emphasized pre-emptive intake isolation, \textit{``we should not wait for the AI to misremember and then fix it in settings. Users should supply contextual boundaries before the AI interprets the utterance.''}

\subsubsection{Design Expectations}

Participants articulated several expectations about the design, where they advocated for context-decoupled, semantically discrete, and dynamically steerable memory misalignment solutions.

\textbf{Decoupling read and write controls.} An interesting finding was that participants rejected the conventional monolithic view of memory management. Currently, conversational agents govern memory through binary, global toggles. However, participants emphasized that reading historical memory and deciding ongoing memory usage are orthogonal decision vectors that demand asymmetric configurations. For example, users often require a ``read-only'' mode, retrieving preferences without polluting the profile with ad-hoc explanatory context, or conversely, a ``write-only'' mode that logs a novel instruction without biasing the current response with legacy priors. As P5 argued, \textit{``using past memories and persisting new ones are two distinct decisions. I might want it to use my historical memory without saving anything from the current conversation. Alternatively, I might want it not to use past context, yet still persist what I say now.''}

\textbf{Direct provenance over complicated linking and reasoning.} In addressing explainability, participants challenged the prevailing design toward comprehensive knowledge graphs. Visualizing associative strengths or internal correlations between disparate pieces of historical evidence was deemed visual clutter that distracts from pragmatic utility. As P1 stated, \textit{``representing the associative strength between two pieces of evidence is largely pointless ... What genuinely matters is how a specific piece of evidence informs memory usage.''} Therefore, interfaces should exclusively surface directed actionable causal paths that link a discrete evidence node directly to the generated output. Crucially, these causal representations must function as actionable control surfaces rather than static readouts, empowering users to prune a faulty premise and trigger downstream re-synthesis.

\textbf{In-situ interruption during the reasoning phase.} Finally, participants identified an unaddressed temporal focus for memory intervention, in-generation steering during the model's reasoning phase. Existing mechanisms primarily focus on pre-task boundary specification or post-hoc output curation. However, because early retrieval of inferential deviations compound as generation unfolds, waiting for complete response synthesis imposes unnecessary cognitive and temporal overhead. P2 underscored the utility of inspecting inference streams in real time, \textit{``when I see it formulate an incorrect premise midway through reasoning, I feel an immediate urge to interrupt it ... Once an intermediate step derails, the entire downstream output is ruined. Being able to interact with and steer that reasoning trajectory in real time would be vastly more efficient.''} Enabling real-time inspection during active inference therefore represents a transformative interaction for preventing error propagation in agentic memory.

\subsubsection{Design Trade-offs}

Across all four workshop sessions, participants identified three tensions related to human-memory interaction, especially on resolving memory misalignment problems.

\textbf{Cognitive overhead vs. autonomous processing.} Proactive confirmation mechanisms offer high fidelity but risk user fatigue. P7 observed, \textit{``I rarely touch memory settings because I simply do not want to spend time managing them. Yet without control, errors feel invasive.''} P7 further recalled a contextual collision where the system awkwardly incorporated leisure hobbies into professional queries, \textit{``I once asked about my research, and it used my piano hobby as an analogy. I felt offended, it felt strained, dragging my personal leisure into a professional settings.''} To resolve this without manual burden, P10 proposed risk-gated prompting, \textit{``the system should only actively prompt when an attribution is ambiguous and the potential error would severely impact downstream usability.''} 

\textbf{Granularity vs. usability.} While detailed provenance offers interpretability, participants warned against excessive complexity. Regarding parametric dials, P5 observed that abstract numerical sliders may confuse user mental models, \textit{``sliders can feel too abstract without concrete semantic values. Users might wonder what a specific retrieval depth actually entails.''} Instead, P8 recommended progressive summarization, \textit{``summarizing into a single concise sentence is best ... Users can click to navigate into the original source dialogue only if they need deeper verification.''}

\textbf{Ephemeral context vs. long-term identity.} A pervasive concern was distinguishing transient task artifacts from durable traits. P5 stressed splitting read and write permissions, \textit{``using past memories and persistent current utterances are two separate decisions. Users often want past context applied without storing the current ephemeral session.''} Corroborating this, P11 proposed a dynamic promotion model where recurring conversational items graduate to long-term memory only after passing a probationary evaluation threshold, preventing ad-hoc queries from corrupting the user's permanent persona.

\section{RQ3: Evaluating User Perceptions of Mitigation Strategies}

Along with the 12 candidate interaction designs, participants in RQ1 also surfaced underlying tensions, such as around supervisory control and cognitive load. How end users navigate these trade-offs and evaluate the efficacy of such mechanisms remains unknown. To investigate these perceptions across such a wide design space, we conducted a speed dating study~\cite{jin2022exploring,xu2026fragmentation} paired with interactive prototypes.

\subsection{Methodology}

\subsubsection{Speed dating design} Following prior work~\cite{jin2022exploring,xu2026fragmentation}, we conducted a speed-dating study to evaluate the 12 designs developed in the workshop. Each participant was randomly assigned three designs, with assignment intended to balance the number of evaluations across designs.

\begin{figure}[!htbp]
    \centering 
    \includegraphics[width=\textwidth]{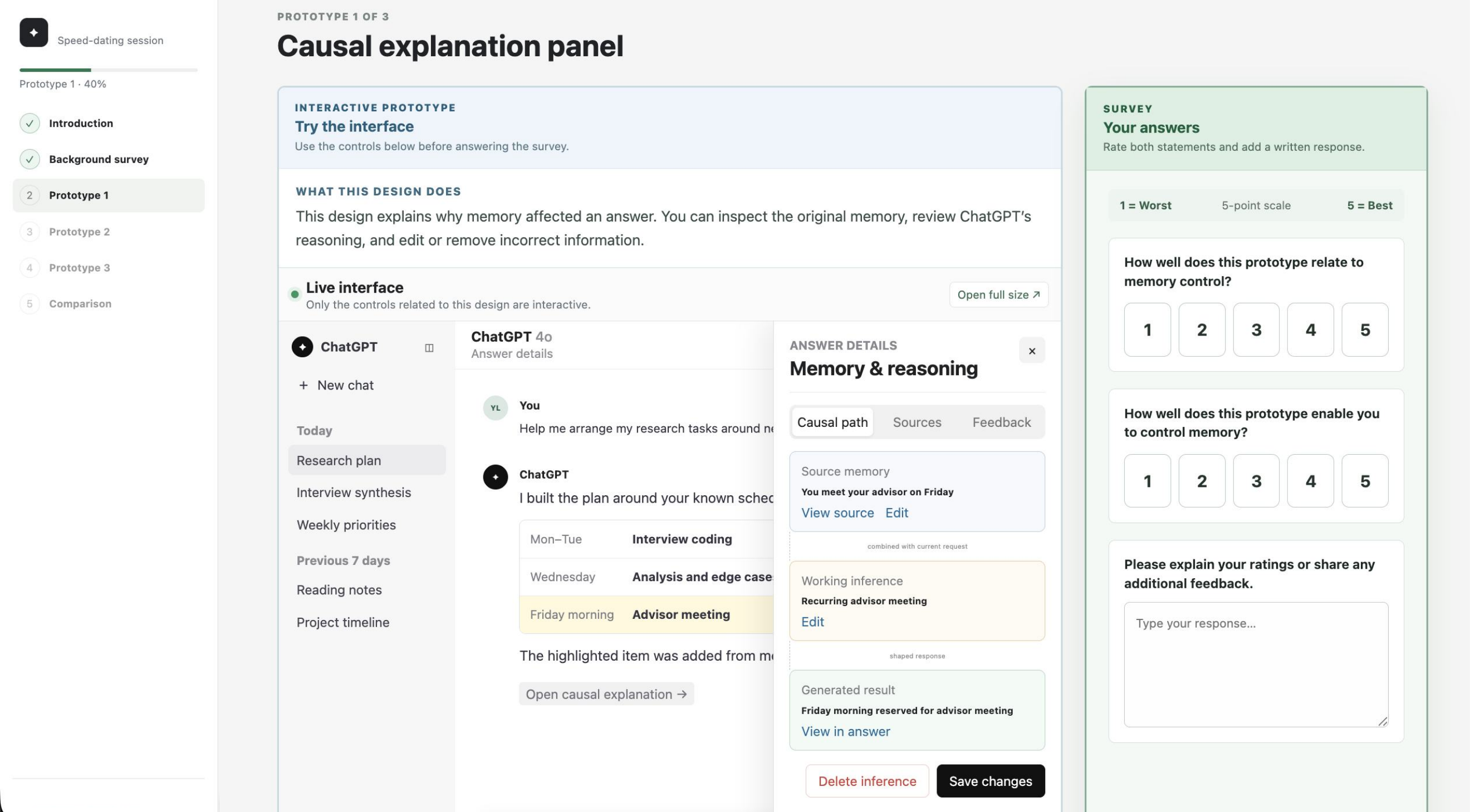}
    \caption{The experiment interface for speed dating sessions.}
    \Description{
    Interface used in the speed dating study. For each sampled design, participants were shown its storyboard together with a high-fidelity interactive prototype that they could explore before evaluating the design. Participants then rated whether the design helped them relate to the memory misalignment problem and whether it could address that problem.
    }
    \label{fig:experiment_interface}
\end{figure}

For each design, we developed a storyboard and a high-fidelity interactive prototype, both available in our \href{https://anonymous.4open.science/r/CHI27-4B3E/}{anonymized supplementary repository}. Participants reviewed the storyboard and could freely explore the corresponding prototype before evaluating the design (Figure~\ref{fig:experiment_interface}). We asked participants to assess two aspects: (i) whether the design helped them relate to the memory misalignment problem and (ii) whether they believed the design could address that problem.

% For experienced person, for example self-identified researchers and designers, we asked:

% $\bullet$ Whether the design is effective? 

% $\bullet$ Whether the design is easy to deploy?

% $\bullet$ Whether the design is cognitively demanding or hard to understand?

\subsubsection{Recruitment and participants.} We recruited participants through Prolific. Eligibility criteria included a Prolific approval rate of at least 95\% and prior experience using LLM-based applications. We targeted 96 participants completing the study, corresponding to 288 design evaluations, or 24 evaluations per design. This target was based on a 12-group omnibus-effect approximation targeting 80\% power at $\alpha=.05$ for a medium effect (Cohen's $f=.25$). To accommodate attrition and prespecified exclusions, we planned to recruit approximately 115--120 participants.

We recruited 121 participants, including 61 men, 56 women, and 4 participants in other gender categories. Participants' ages ranged from 19 to 82 years (median $=36$). Most participants identified as White ($n=67$, 55.4\%), followed by Black ($n=25$, 20.7\%) and Asian ($n=14$, 11.6\%). Most resided in the United Kingdom ($n=88$) or the United States ($n=31$). Detailed demographics are reported in Table~\ref{tab:memory-buddy-participants}. Participants were compensated at GBP~9 per hour. The study was approved by our institution's Institutional Review Board (IRB).

\subsubsection{Data analysis} For quantitative ratings, we modeled design strategy as a fixed effect and included a participant-specific random intercept to account for repeated evaluations. Significant main effects were followed by post-hoc pairwise comparisons with Holm--Bonferroni correction.

Two authors analyzed the qualitative responses using a hybrid deductive--inductive coding approach. Deductive coding drew on the design dimensions and memory misalignment dimensions, while additional codes were developed inductively from participants' responses. The authors subsequently organized the emergent codes into subthemes and themes. We did not calculate inter-rater reliability; instead, we supported analytic consistency through periodic discussions of coded excerpts and interpretations~\cite{mcdonald2019reliability}.

\subsection{Results}

\subsubsection{Quantitative Feedback}

\begin{figure}[!htbp]
    \subfloat[Memory ranking preferences.]{
        \includegraphics[width=0.6\textwidth]{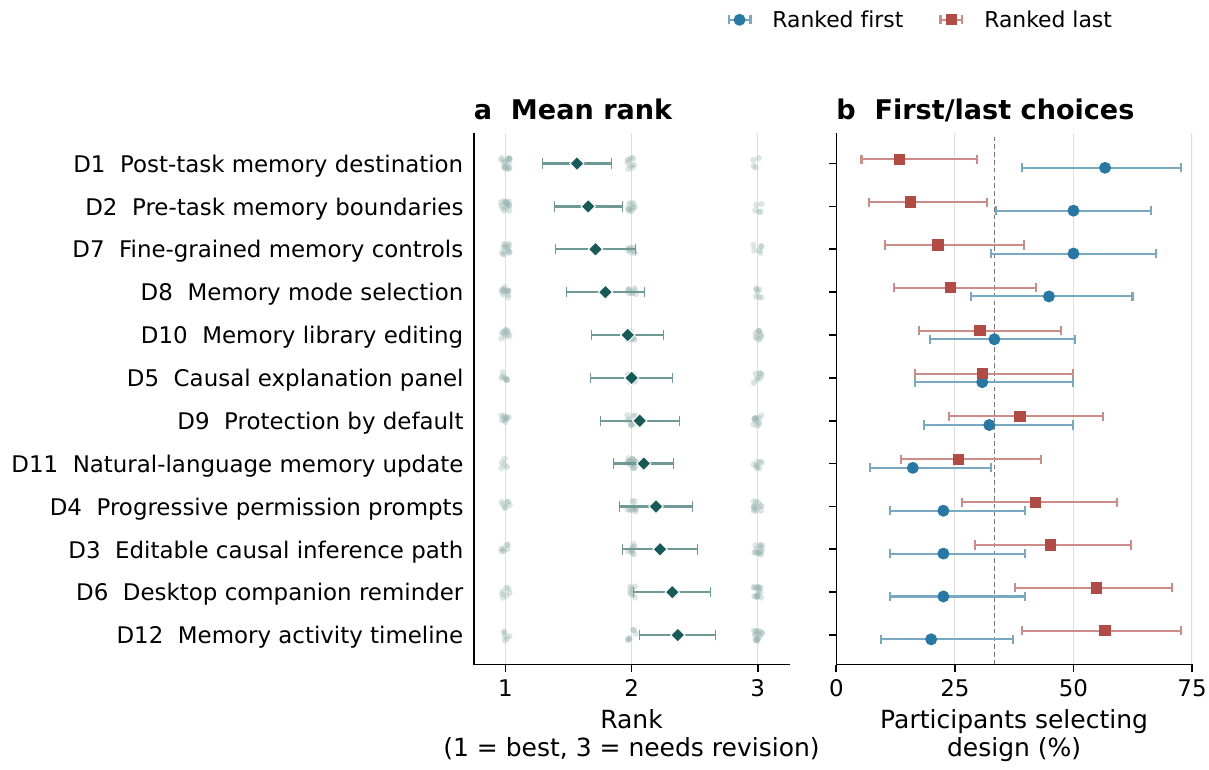}
    }
    \subfloat[Ranking priority.]{
        \includegraphics[width=0.4\textwidth]{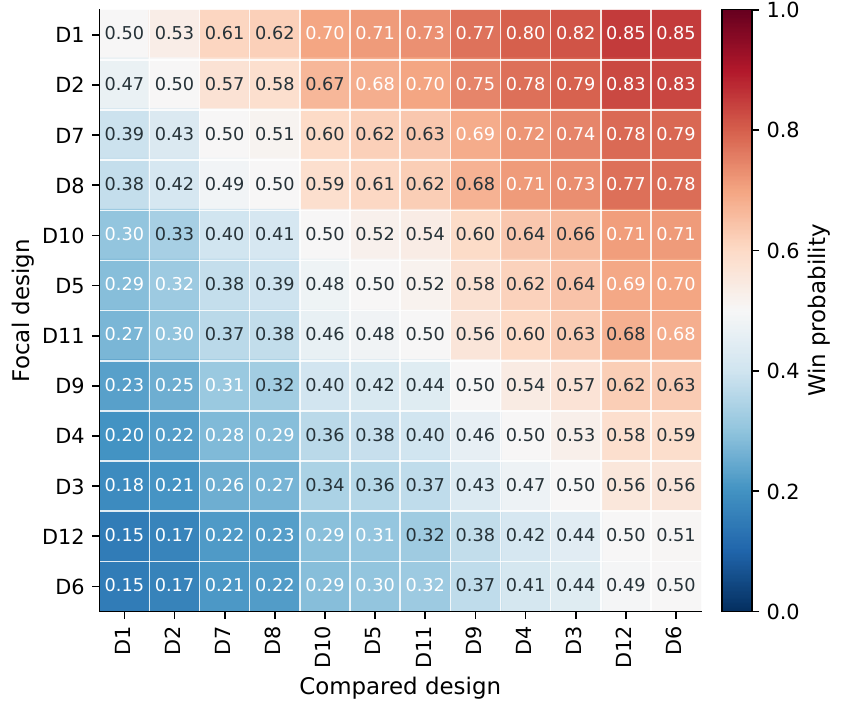}
    }
    \caption{(a) Memory ranking preferences, and (b) priority of ranking.}
    \Description{
    Ranking results for the twelve memory designs. Panel (a) shows each design's mean rank, with rank 1 indicating the most preferred and rank 3 the least preferred, together with the proportions of participants who ranked each design first or last. Post-task memory destination, pre-task memory boundaries, and fine-grained memory controls were among the most preferred designs, whereas the desktop companion reminder and memory activity timeline ranked relatively low. Panel (b) shows pairwise ranking priorities between designs as a probability heatmap, broadly reflecting the same preference ordering.
    }
    \label{fig:ranking}
\end{figure}

\begin{figure}[!htbp]
    \subfloat[Problem relation.]{
        \includegraphics[width=0.5\textwidth]{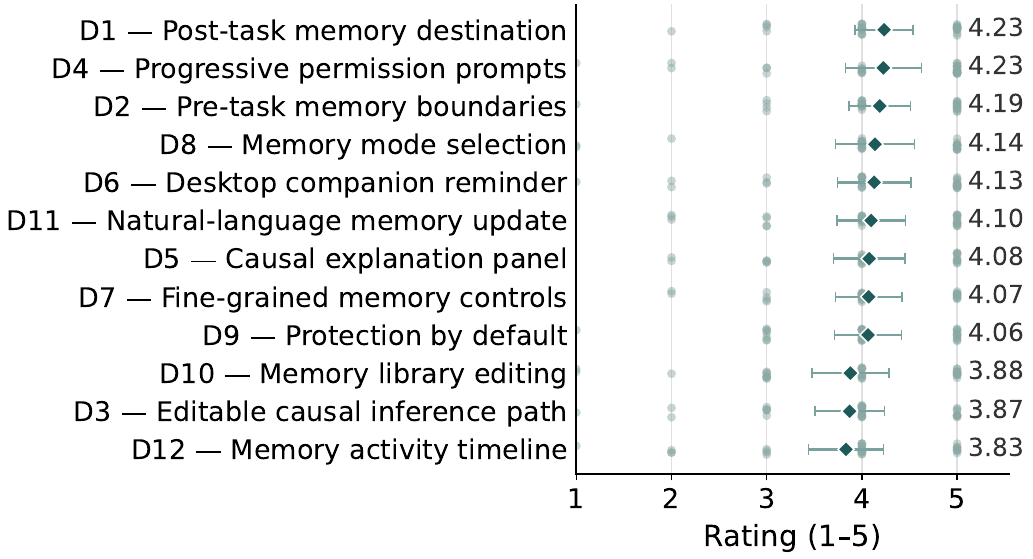}
    }
    \subfloat[Memory control.]{
        \includegraphics[width=0.5\textwidth]{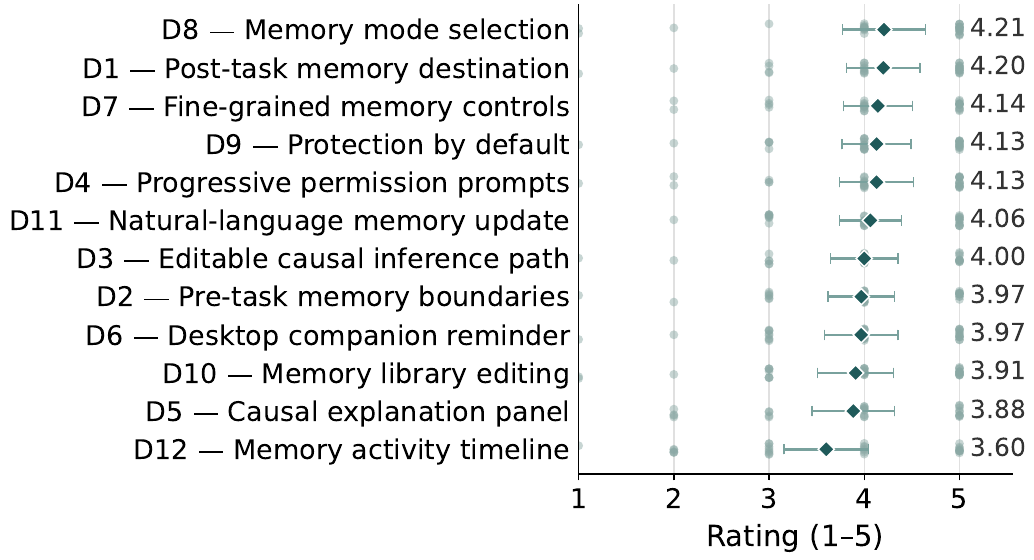}
    }
    \caption{(a) Whether the participants could relate to the problem, and (b) whether the participants thought the design exerted control. Errorbar indicates 95\% confidence interval.}
    \Description{
    Mean ratings for the twelve memory designs on two five-point measures, with error bars showing 95\% confidence intervals. Panel (a) reports how well participants could relate to the problem represented by each design; ratings were generally high, ranging from approximately 3.8 to 4.2. Panel (b) reports perceived memory control; most designs received ratings around 4, with memory mode selection and post-task memory destination rated highest and the memory activity timeline rated lowest.
    }
    \label{fig:rating}
\end{figure}

The final forced-choice ranking provided evidence about relative preference, where we found significant effects of designs on preferences ($p<.001$, 100,000 permutations). Descriptively, D1, D2, D7, and D8 received the highest mean ranking positions, whereas D6 and D12 were most frequently placed last. 

Because each participant evaluated only three of the twelve designs, the rating data were analyzed using a participant-stratified randomization procedure. For each outcome, the three design labels assigned to a participant were permuted within participant while the observed ratings were held fixed, thereby preserving the incomplete-block structure and within-participant dependence. The omnibus design effect was not statistically reliable for perceived relevance to memory control ($p=.679$) or perceived control over memory ($p=.302$).

The Plackett–Luce analysis indicated reliable heterogeneity in participants’ design preferences, likelihood-ratio ($\chi^2_{11}=38.04$, $p<0.001$). Post-task memory destination (D1) received the highest preference weight (0.183), followed by pre-task memory boundaries (D2, 0.159), fine-grained memory controls (D7, 0.119) and memory mode selection (D8, 0.113). In contrast, desktop companion reminder (D6, 0.032) and memory activity timeline (D12, 0.033) received the lowest weights. The matrix in Figure~\ref{fig:ranking} visualizes these differences as pairwise win probabilities. For example, the estimated probability that D1 would be preferred to D6 was approximately 0.85, while the probability that D6 would be preferred to D1 was approximately 0.15. Plackett–Luce worth estimates were moderately correlated with mean relation ratings ($\rho=.46$, $p=.131$) and perceived memory control ratings ($\rho=.48$, $p=.114$), suggesting that overall ranking preferences were related to the Likert-scale evaluations.

\begin{figure}[!htbp]
    \centering
    \includegraphics[width=0.6\textwidth]{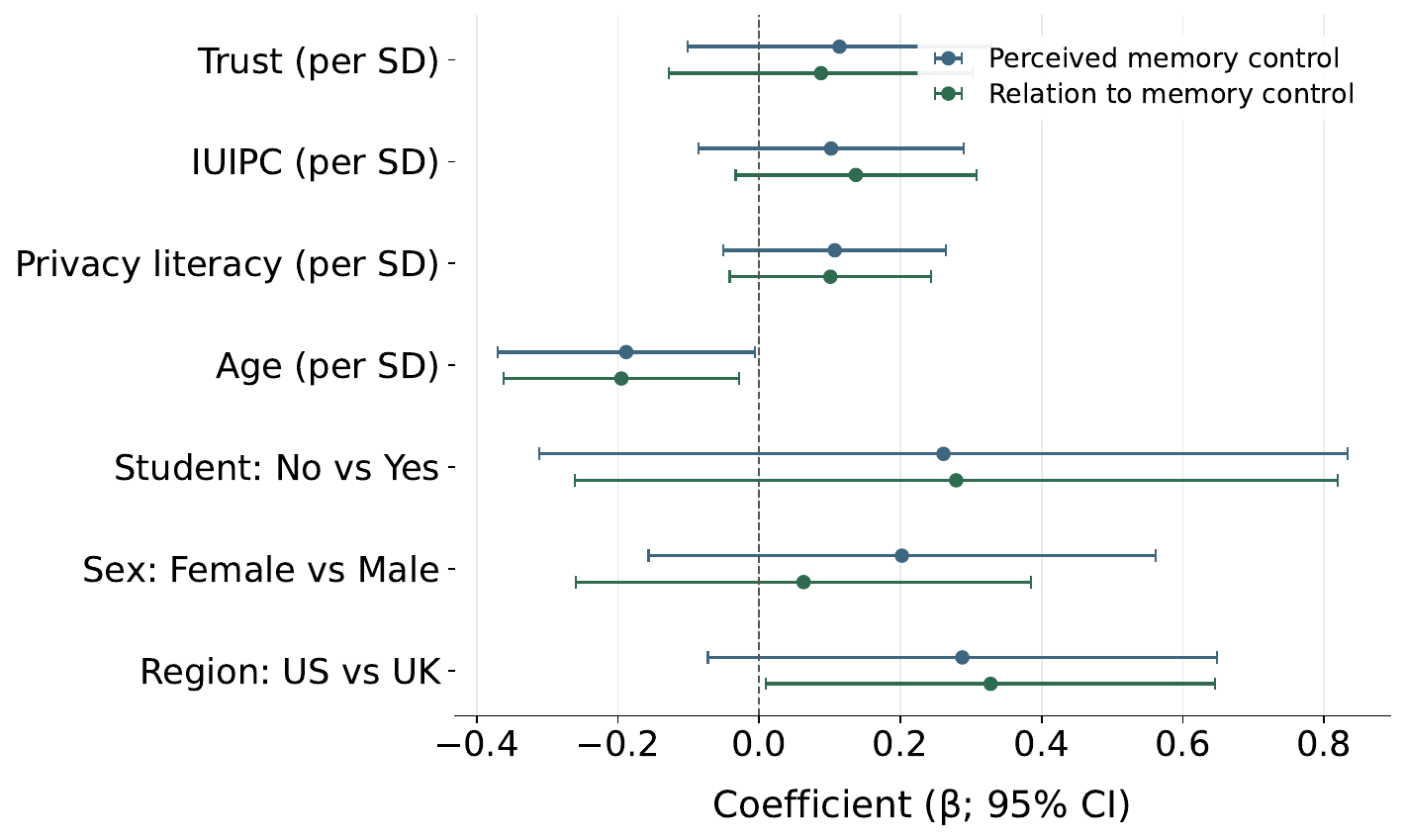}
    \caption{The regression of ratings on demographic and individual-related variables.}
    
    \Description{
    Forest plot showing regression coefficients and 95\% confidence intervals for demographic and individual-level predictors of participant ratings. Points represent estimated coefficients and horizontal lines represent confidence intervals, with the vertical zero line indicating no association. Older age was associated with lower ratings, while participants from the United States gave higher ratings than those from the United Kingdom. Other measured individual factors, including privacy literacy, IUIPC, and trust in AI, showed no statistically significant associations.
    }
 
\label{fig:demographics}
\end{figure}

As shown in Figure~\ref{fieg:demographics}, for relation to memory control, older age was significantly associated with lower ratings ($\beta=-0.20$, [-0.36, -0.03], $p=.022$), whereas participants from the United States gave significantly higher ratings than those from the United Kingdom ($\beta=0.33$, [0.01, 0.65], $p=.043$). Privacy literacy ($\beta=0.10$, [-0.04, 0.24], $p=.165$), IUIPC ($\beta=0.14$, [-0.03, 0.31], $p=.115$), and trust in AI ($\beta=0.09$, [-0.13, 0.30], $p=.425$) were positively associated with ratings, but none of these associations reached significance. Neither sex nor student status was significantly associated with perceived memory control.

A similar pattern was observed for perceived memory control. Older age was again significantly associated with lower ratings ($\beta=-0.19$, [-0.37, -0.01], $p=.043$). The coefficients for privacy literacy ($\beta=0.11$, [-0.05, 0.26], $p=.182$), IUIPC ($\beta=0.10$, [-0.09, 0.29], $p=.287$), and trust in AI ($\beta=0.11$, [-0.10, 0.33], $p=.300$) were positive but not significant. The difference between participants from the United States and the United Kingdom was also positive but non-significant for perceived memory control ($\beta=0.29$, [-0.07, 0.65], $p=.117$). Sex and student status were not significantly associated with this rating dimension. Overall, age was the only predictor that showed a significant association with both rating outcomes, while the effects of privacy literacy, IUIPC, trust in AI, sex, and student status were not distinguishable from zero in either model. 

\subsubsection{Qualitative Feedback}

We analyzed participants' feedback using a hybrid strategy, emphasizing three pre-defined dimensions while seeking new undisclosed dimensions. Participant identifiers are pseudonymised as P1--P121.

\textbf{\textit{Interaction Form}.} Participants broadly preferred panel-based interaction because it converted abstract memory states into inspectable diagnostic interventions. As P25 explained, \textit{``I can review the suggested memory before it is saved ... edit it, and choose to save it for future chats, use it only in this chat, or not save it.''} P16 similarly valued how direct manipulation provides \textit{``a surgical level of control that prevents unwanted behavior without requiring a full memory wipe.''} Avatar-based designs fostered timely ambient awareness. P08 noted the design \textit{``let me check and review new memories the second they happen''}. Conversational updates also offered intuitive natural language negotiation with visual diffs (P63). However, conversational modes introduced ambiguity regarding final system states (P84). Across forms, participants valued category-based granularity (P06, P09) and causal explanations linking outputs to source memories (P06, P07). Crucially, however, they distinguished passive transparency from actionable control. As P28 remarked of the activity timeline, \textit{``it feels more like a monitoring tool than a true control interface,''} without direct editing affordances.

\textbf{\textit{Placement}} Participants' preferences regarding placement revealed a direct tension between situational context and workflow continuity. In-chat mechanisms were praised for situating memory decisions directly at the point of action. For example, P19 appreciated granular per-memory permissions prior to output generation. However, this coupling introduced severe interaction overhead when invoked repeatedly. P05 and P32 cautioned that prompt fatigue could quickly \textit{``delay the ease of use of ChatGPT''} when processing multiple candidate memories. Conversely, out-of-chat placements such as dedicated sidebars or dashboards supported deliberate, retrospective auditing without cluttering the active dialogue stream (P01, P16). Yet, separate inspection spaces risked detaching visibility from direct intervention, leaving users frustrated when monitoring mechanisms lacked immediate operational layers (P12, P28).

\textbf{\textit{Intrusiveness}} Participants emphasized that the acceptable level of intrusiveness depends on task criticality and perceived risk. Low-intrusiveness designs featuring default protection or passive logging preserved conversations, with participants appreciating systems that automatically withhold sensitive data by default (P14, P32). Nonetheless, overly unobtrusive mechanisms induced skepticism regarding whether protection was reliably executed without explicit verification (P22, P26). At the other extreme, high-intrusiveness causal inspections offered important accountability for consequential reasoning (P06), but participants overwhelmingly rejected them as default interfaces due to the cumulative cognitive burden of continuous verification (P32, P78). Therefore, those designs that are not overly intrusive emerged are the most preferred, as participants thought structured manageable checkpoints without derailing interaction flow (P10, P14).

\textbf{\textit{User agency and contextual boundaries.}} Across all evaluated dimensions, participants emphasized that memory transparency is functional only when coupled with actionable control and contextual boundaries. Users explicitly distinguished passive observation from actionable intervention. As P28 observed, visibility without modification mechanism yields \textit{``more like a monitoring tool than a true control interface''}, while inspecting source memories could build trust by enabling users to audit and discard incorrect inputs (P06, P16, P19). Crucially, participants favored fine-grained boundary management than uniform system defaults, articulating that memory controls must decouple and independently govern distinct life domains such as professional work, personal hobbies, and sensitive disclosures (P02, P09).

% \textbf{\textit{Control versus awareness.}} Across all three dimensions, participants distinguished between being informed about memory activity and being able to change it. P28 said that \textit{``The timeline and searchable history make the process more transparent. However, it feels more like a monitoring tool than a true control interface.''} In contrast, P19 valued designs, as \textit{``I like that I can see where the memory came from and decide whether to keep or delete it.''}. 

% \textbf{\textit{Transparency as a prerequisite for control.}} Participants generally treated transparency as valuable because it supported decision-making. P16 said that tracing where and when a memory was inferred builds trust. P06 similarly valued a design that showed the source memory and made it possible to identify and remove incorrect input. 

% \textbf{\textit{Personalisation and contextual boundaries.}}
% Participants wanted memory controls to reflect differences between personal, work-related, scheduling, and sensitive information. P02 said that users should be able to ``select the different categories that the system remembers (and where), so there's plenty of flexibility.'' P09 said, ``I like how you can specialise it to specific preferences.'' The corresponding codes were category-level control, context-specific retention, sensitive-information protection, and user-defined defaults.

\section{Discussions}

Across empirical and design inquiries, this paper characterized memory misalignment phenomena (RQ1), conceptualized a mitigation design space across form, placement and intrusiveness (RQ2), and evaluated user trade-offs across candidate mechanisms (RQ3). Synthesizing these findings, we contribute an interactional breakdown based perspective on memory misalignment, and in the next subsections we discuss how these insights could advance human-centered memory design.

\subsection{Advance AI's Memory with Human-in-the-loop}

We delineate how our findings advance agentic memory across four dimensions from theoretical contributions to empirical insights: establishing the first human-centric memory misalignment taxonomy, supplying localized supervisory signals for dynamic AI alignment, deepening AI community's understanding for memory alignment, and providing interaction guidelines for HCI system design.

\textbf{Establishing a human-centric memory misalignment taxonomy.} A core contribution of our work is providing a novel human-centric perspective on AI's memory, which differs from existing taxonomies that focuses on operational lifecycles~\cite{jones2025storage,yang2026toward}, representations~\cite{hu2025memory,wu2025human,zhang2025survey}, temporal and functional scopes~\cite{du2026memory,huang2026survey}, and cognitive tiers~\cite{luo2026storage}. Our work extends their assumption of faithful information retention, transcending how memory should function to investigate how memory misaligns with human expectations. We also bridge the investigations on human memory perception~\cite{chen2026relational}, and those human-like memory architectures~\cite{honda2025human,westhausser2026caim} by mapping system failures across the agent lifecycle to human alignment preferences and cognitive memory errors (e.g., Schacter's memory sins~\cite{schacter1999seven}). 

\textbf{Supplying supervisory signals for dynamic AI alignment.} For the broader AI and machine learning communities, our taxonomy shifts memory evaluation from end-to-end performance benchmarks toward fine-grained interpretable diagnostic signals. Contemporary personalization models struggle to diagnose why retrieved context degrades response utility, sometimes conflating surface-level retrieval failure with underlying semantic mismatch~\cite{balepur2025good,xiong2026memory}. By pinpointing specific failure mechanisms, such as distinguishing between temporal staleness in storage and over-application during retrieval, our taxonomy provides the granularity required for targeted intervention. Rather than relying on global fine-tuning or prompt patching, AI developers can operationalize these failure types as evaluation metrics~\cite{zhang2026characterizing}, negative reward signals in preference optimization~\cite{abdolmaleki2025learning}, or dynamic gating heuristics~\cite{han2024efficient}. This enables agents to treat real-time human feedback as structured corrections.

\textbf{Deepening AI community's understanding for memory alignment.} By integrating human-in-the-loop oversight into agent memory, this work advances the paradigm of agentic personalization from static, offline preference optimization~\cite{ji2023ai,zhong2024panacea,li2026lifealign} to interactive co-alignment~\cite{arzberger2026co,zhang2025align}. While conventional alignment frameworks implicitly assume uncorrupted context ingestion~\cite{xu2026single} and reduce behavioral breakdowns to algorithmic hallucinations~\cite{chen2025halumem}, our empirical findings reveal that memory breakdowns are sometimes relational, originating from violated contextual boundaries during situated interactions. Therefore, we advocate that resolving these tensions requires reconsidering the human user as an active co-regulator of memory states~\cite{fan2025user}.

\textbf{Providing interaction guidelines for HCI system design.} For interactive design, our work bridges the divide between passive error categorization and mitigation interfaces. Prior human-in-the-loop memory systems frequently default to rich data exposure, such as inspectable knowledge graphs or record tables~\cite{huang2023memory,vaithilingam2025semantic,yen2024memolet}, which may burden users with cognitive frictions when tackling misalignments in real usage~\cite{shirali2026burden}. Extending this work, we translate our 14 failure modes into 12 mitigation strategies structured across a three-dimensional design space. By showing that users prefer lightweight, proactive verification at natural task boundaries over cognitively exhausting causal retrospection, our findings deliver empirical guidelines for balancing supervisory agency against interaction overhead. 

\subsection{Design Implications}

Our findings indicate that designing agent memory requires addressing three challenges spanning the memory lifecycle, including its generation, application and interaction. We outline these design implications grounded in our taxonomy. 

\textbf{Generation: Context decoupling and granular decay of memory.} Algorithmic memory implementations often suffer from a unilateral aggregation dilemma, where conversational fragments are indiscriminately ingested and aggregated into monolithic user profiles~\cite{dash2026algorithmic,chen2026relational}. Our taxonomy reveals that design flaws are the primary catalyst for temporal staleness and outdated contexts. In line with cognitive inspired methods like ACT-R~\cite{honda2025human,westhausser2026caim}, memory mechanisms must decouple transient episodic interactions from semantic identities. Systems should support multi-tier decay mechanics and explicit abstraction boundaries. By contextualizing memory validity windows and adopting privacy-by-default heuristics, agents can prevent exploratory one-off inquiries from polluting core behavioral representations, mitigating the unforgetful and invasive behavior observed in contemporary chatbots~\cite{chen2026relational,jones2025users}.

\textbf{Application: Context-sensitive grounding.} Contemporary retrieval-augmented generation (RAG) and long-term memory architectures rely heavily on vector similarity to recall relevant persona context~\cite{wu2025long,xu2026single}. However, this design conflates semantic proximity with pragmatic applicability. As evidenced by our taxonomy, a large proportion of usage misalignments, specifically mismatch and overly coarse memory use, manifest when agents rigidly project accurate historical traits into incongruent interactional contexts. This corroborates Balepur et al.'s observation that surface-level similarity often divorces retrieval from genuine utility~\cite{balepur2005good}, as well as Cheng et al.'s critique of uncalibrated agent assumptions~\cite{cheng2026verbalizing}. To resolve this dissonance, memory retrieval must evolve from an opaque, deterministic pipeline into a bidirectional grounding process~\cite{arzberger2026co,li2026alignment}. Rather than binding retrieved memories into prompt context, agent architectures should incorporate dynamic context applicability filters that evaluate not only what to retrieve, but whether the target context warrants the application of historical persona priors. When contextual ambiguity or stylistic mismatches arise, systems should externalize model rationale through lightweight verification prompts, or conversational natural language steering. Treating memory application as an ongoing dialogue act calibrates mutual expectations and prevents context-blind preference enforcement~\cite{rahman2026vibe,zhang2025align}.

\textbf{Interaction: Rebalancing control against cognitive overhead.} Prior human-in-the-loop systems often address agency deficits by treating memories as explicit, exposed data artifacts, such as \textit{Memolet}~\cite{yen2024memolet}, \textit{Memory Sandbox}~\cite{huang2023memory}, and \textit{Semantic Commit}~\cite{vaithilingam2025semantic}. While providing deep oversight, our speed dating evaluations highlight a design trade-off, where high-intrusiveness, cognitively demanding interventions were perceived as burdensome and received lower control ratings. This resonates with Shirali's formalization of alignment as an interactive burden~\cite{shirali2026burden}, where excessive operational friction degrades user welfare. Conversely, participants expressed preferences for lightweight interventions embedded at natural task boundaries. This suggests that agent architectures should adopt a friction-aware paradigm. Rather than exposing complete reasoning chains by default, systems should operate under a low-friction method, using proactive boundary prompts and post-task summaries while reserving detailed causal inspection.

\section{Ethical Considerations}

Our research was guided by the ethical principles outlined in the Belmont Report~\cite{beauchamp2008belmont} and Menlo Report~\cite{bailey2012menlo}, structured around four aspects that these two reports mentioned. Notably, our research received approval from our university's Institutional Review Board (IRB).

\textbf{Respect for persons.} All phases of the study, including the memory export collection, the diary study, the co-design workshops, and the speed-dating evaluation, were formally reviewed and approved by our institution's IRB. All participants provided informed consent after being briefed on the research objectives, procedures, potential risks, and their right to withdraw. Participants in the empirical study has the right to withhold, redact or decline to report their memory data.

\textbf{Beneficence.} Under the beneficence principle, we sought to maximize research utility while minimizing physical, psychological, and privacy risks. Adhering to data minimization, researchers refrained from capturing or inspecting participants' raw, longitudinal chat logs. All collected data were anonymized for PII, and assigned unique pseudonyms (e.g., P1). Research records were stored exclusively in access-restricted repositories accessible solely to the primary research team.

\textbf{Justice.} The justice principle requires fair distribution of research burdens and benefits, precluding exploitative practices and biased exclusion. Participant recruitment across all phases targeted diverse demographics, occupational sectors, and technical literacy levels. Participants also received fair and equitable compensation. Diary study participants were compensated based on their participation (a base of 20 CNY plus 1 CNY per entry, up to 200 entries). This criteria is set according to the local wage standard. Participants in the workshop study was compensated 300CNY according to the local wage standard. Participants in the speed dating study were compensated at a rate of 9GBP/hour, meeting Prolific's recommended compensation standards.

\textbf{Respect for law and public interest.} Reflecting the Menlo Report's guidance on societal responsibility, compliance and dual-use considerations, this research explicitly aims to advance public interest by exposing how conversational agents can cause friction in user interactions. Rather than promoting in-comfort profiling, we advocate for blueprints on privacy-by-default and aligned memory designs.

\section{Limitations}

We acknowledge several limitations in this paper. First, we did not manually examine participants' chat dialogues due to privacy concerns. Therefore, we did not verify ground truths when constructing the taxonomy. Although we asked participants to report faithfully, and told them report wrongly may result in reduced compensation, there are still risks of participants' cheating, and misunderstandings. Second, our participants are mostly Chinese users, designers and researchers, and those participants are subject to recruitment bias. However, China is a representative region where LLMs are developing rapidly (e.g., with Qwen, Deepseek and Kimi models), and users' adoption of LLMs are relatively prevalent. Nevertheless, future work should broaden the scope and cross-validate memory misalignment cases in other cultural and regional contexts. Third, we primarily examined the memories of LLM-based chatbots such as ChatGPT, while acknowledging there are also memories in agentic systems such as Claude code, Codex, whose misalignment issues warrant future work.

\section{Conclusion}

Memory mechanisms are integral to personalization in LLM-based conversational agents. However, memory misalignment, where generation and usage of memory does not align with users' expectations, are understudied. In this paper, we conducted three studies to characterize, address and evaluate memory misalignment from user perspectives. Grounded in cognitive memory error frameworks, our empirical investigation of real-world memory reports and dairies formed a taxonomy of 14 misalignment types spanning intake, storage and management, and retrieval and interpretation stages. Through co-design workshops with experienced HCI researchers, we derived 12 interaction strategies organized along three dimensions: interaction form, placement, and intrusiveness. Subsequent evaluations with 121 users via speed dating and Plackett-Luce modeling showed a strong preference for proactive, lightweight controls, such as post-task destination prompts and pre-task boundary settings, over cognitively demanding casual inspections or passive activity logging. Our taxonomy, design space and empirical trade-offs provide important base for designing aligned memory systems.

% \begin{acks}
% \end{acks}

% \section*{Ethics and Privacy Statement}

% This section of your ACM work should discuss the potential societal
% risks that might result from its publication; two to three sentences
% related to the findings of your study, or new advancements made
% possible by their developed methods. The privacy and ethics statement
% should clearly address the broader impacts of their work as it relates
% to the authors' interpretation of privacy, fairness, safety, human
% rights, data sovereignty, or future misuse and any benefit/risk
% trade-off resulting from this research. We acknowledge that some
% papers may have minimal societal risks beyond those considered by
% institutional review boards, and the dimensions considered by any
% review of the user study design or dataset licenses could be provided
% in this statement.

\bibliographystyle{ACM-Reference-Format}
\bibliography{sample-base}

\appendix

\section{Generative AI Usage}

We used ChatGPT and Gemini-3.1-pro for polishing the text of this paper, including checking grammar. We also used ChatGPT for refining the figures. All authors hold full responsibility for all this paper's content.

\section{Demographics for Workshops}

Table~\ref{tab:demographics} shows the demographics for the workshops. 

\begin{table}[!htbp]
\centering
\caption{Participant demographics.}
\label{tab:demographics}
\begin{tabular}{cccccc}
\toprule
\textbf{ID} & \textbf{Age} & \textbf{Gender} & \textbf{Current Country of Residence} & \textbf{Research Experience (Years)} & \textbf{Publication Numbers} \\
\midrule
1  & 18-25 & Male   & China & 6 & 20+ \\
2  & 26-35 & Female & United States & 3 & 3 \\
3  & 26-35 & Female & Australia & 7 & 10+ \\
4  & 26-35 & Male & Japan & 10 & 10+ \\
5  & 26-35 & Female & United States & 3 & 3 \\
6  & 18-25 & Female & United States & 3 & 1 \\
7  & 18-25 & Female & China & 3 & 2 \\
8  & 26-35 & Female & Australia & 6 & 10+ \\
9  & 26-35 & Male & China & 3 & 3 \\
10 & 26-35 & Female & China & 5 & 7 \\
11 & 26-35 & Female & United States & 3 & 3 \\
12 & 26-35 & Female & Finland & 3 & 3 \\
\bottomrule
\end{tabular}
\end{table}

\section{Demographics for Speed Dating}

Table~\ref{tab:memory-buddy-participants} showed the participants' demographics in the speed dating study.

\begin{table}[!htbp]
\centering
\caption{Participant demographics in the speed dating study (N = 121).}
\label{tab:memory-buddy-participants}
\small
\begin{tabular}{llrr}
\toprule
\textbf{Variable} & \textbf{Category} & \textbf{n} & \textbf{\%} \\
\midrule
Age & 18--24 & 11 & 9.1 \\
    & 25--34 & 34 & 28.1 \\
    & 35--44 & 34 & 28.1 \\
    & 45--54 & 21 & 17.4 \\
    & 55--64 & 15 & 12.4 \\
    & 65+ & 4 & 3.3 \\
    & Prefer not to disclose & 2 & 1.7 \\
Sex & Male & 61 & 50.4 \\
    & Female & 56 & 46.3 \\
    & Prefer not to say & 4 & 3.3 \\
Ethnicity & White & 67 & 55.4 \\
    & Black & 25 & 20.7 \\
    & Asian & 14 & 11.6 \\
    & Mixed & 7 & 5.8 \\
    & Other & 3 & 2.5 \\
    & Prefer not to say & 5 & 4.1 \\
Country of residence & United Kingdom & 88 & 72.7 \\
    & United States & 31 & 25.6 \\
    & Missing & 2 & 1.7 \\
Student status & No & 93 & 76.9 \\
    & Yes & 13 & 10.7 \\
    & Missing & 15 & 12.4 \\
Employment status & Full-time & 58 & 47.9 \\
    & Part-time & 17 & 14.0 \\
    & Unemployed and seeking work & 14 & 11.6 \\
    & Not in paid work & 13 & 10.7 \\
    & Due to start a new job & 2 & 1.7 \\
    & Other & 2 & 1.7 \\
    & Missing & 15 & 12.4 \\
\bottomrule
\end{tabular}
\end{table}

\section{Experiment Materials}

This appendix describes the materials used in the two stages of the experiment: the Memory Export phase and the Workshop phase. The materials were provided to participants in sequence. In the first phase, participants elicited and evaluated the memories retained by their LLM-based chatbots. In the second phase, participants reviewed a taxonomy of memory misalignment and developed design proposals to address such problems.

\subsection{Memory Export Phase}

The purpose of the Memory Export phase was to elicit the memories and user-related context retained by participants' LLM-based chatbots. Participants used the chatbot that they regularly interacted with and entered the following prompt. The prompt was adapted from Anthropic's official memory-import workflow \cite{anthropic2026memoryimport}.

\begin{quote}
I'm moving to another service and need to export my data. List every memory you have stored about me, as well as any context you've learned about me from past conversations. Output everything in a single code block so I can easily copy it. Format each entry as: \texttt{[date saved, if available] --- memory content.}

Make sure to cover all of the following---preserve my words verbatim where possible: Instructions I've given you about how to respond (tone, format, style, ``always do X,'' ``never do Y''). Personal details: name, location, job, family, interests. Projects, goals, and recurring topics. Tools, languages, and frameworks I use. Preferences and corrections I've made to your behavior. Any other stored context not covered above. Do not summarize, group, or omit any entries. After the code block, confirm whether that is the complete set or if any remain.
\end{quote}

After receiving the chatbot's response, participants reviewed the elicited memories individually. For each memory, they indicated whether it was appropriate. When a memory was considered inappropriate, participants explained the reason for their assessment and indicated whether they wished to modify it. If they wished to modify the memory, they described the desired change and explained the rationale for the proposed modification.

\subsection{Workshop Phase}

The Workshop phase investigated how users might address memory misalignment in LLM-based chat assistants through interface and interaction design. Before the workshop, participants received a bilingual workshop document introducing the concept of memory misalignment, the taxonomy described in ~\ref{fig:misalignment}, representative examples for each category, and the workshop design task. Participants were asked to review the taxonomy and examples before developing their designs, so that they shared a common vocabulary for discussing memory-related problems. Participants were asked to use the taxonomy and examples as a common conceptual background for their design work.

The workshop task was described as follows:

\begin{quote}
Before the workshop, please prepare two designs to help users address potential memory mismatch issues in LLM-based chat assistants. It is recommended that you present your designs using a combination of diagrams and text.
\end{quote}

The workshop document also included two illustrative design examples. The first example, ``Dynamic Memory Cards and Active Clarification,'' addressed errors occurring during memory intake and storage. It proposed displaying a pending memory card when the system identified information that might be stored. Users could confirm, edit, delete, or adjust the privacy settings of the proposed memory. The design also included an active clarification mechanism that would ask users whether repeated behaviors represented a stable preference or only a temporary interest. In addition, users could confirm changes to outdated memories and prevent sensitive information from being stored by default.

The second example, ``Memory Transparency Display and Causal Explanation,'' addressed errors occurring during memory retrieval and application. It proposed a collapsible memory panel showing the memories used to generate the current response. Each memory could be linked to its originating conversation, and a causal explanation could describe how the memory influenced the response. Users could edit or delete a memory, add a relevant memory that had not been used, regenerate the response after modifying a memory, or instruct the system to ignore a memory in the current context.

These two designs were provided as illustrative examples of possible design directions. Participants were instructed to replace them with their own designs rather than reproduce the examples directly.

\end{document}